\documentstyle[11pt,aas2pp4]{article}
\received{}
\revised{}
\accepted{}
\slugcomment{}
\lefthead{GORHAM, P.W.}
\righthead{ON RADAR DETECTION OF HIGH ENERGY COSMIC RAYS}
\begin{document}
\title{On the possibility of radar echo detection of ultra-high energy \\ 
cosmic ray- and neutrino-induced extensive air showers}
\author{ Peter~W.~Gorham}

\affil{Jet Propulsion Laboratory, Calif. Inst. of Technology \\
	4800 Oak Grove, Drive, Pasadena, CA, 91109 USA}


\begin{abstract}
We revisit and extend the analysis supporting a 60 year-old suggestion 
that cosmic rays air showers resulting from primary particles
with energies above $10^{18}$ eV should be straightforward to detect
with radar ranging techniques, where the radar echoes are produced
by scattering from the column of ionized air produced by the shower. 
The idea has remained curiously untested since it was proposed, but
if our analysis is correct, such techniques could
provide a significant alternative approach to
air shower detection in a standalone array with high duty cycle,
and might provide highly complementary measurements of air showers
detected in existing and planned ground arrays such as the Fly's Eye or the
Auger Project. The method should be particularly sensitive
to showers that are transverse to and relatively distant from the
detector, and is thus effective in characterizing 
penetrating horizontal showers such as those that 
might be induced by ultra-high
energy neutrino primaries.
\end{abstract}
\keywords{Cosmic-rays, Cosmic-ray detectors, Extensive air showers, 
Neutrinos, Radar}

\newpage

\section{Introduction}

Extensive air showers (EAS) resulting from primary cosmic-ray 
particles of energies 
above 1 EeV ($10^{18}$ eV) produce
an ionization trail which is comparable to that of micro-meteors,
which have been detected for many decades using radar methods
(Lovell 1948; Greenhow, 1952; Hanbury Brown \& Lovell 1962). 
In fact, Blackett \& Lovell (1940) proposed that earlier
detections of sporadic radio reflections at altitudes of 
5-20 km from the
troposphere and mesosphere (cf. Colwell \& Friend 1936;
Watson Watt et al. 1936a, 1936b, 1937; Appleton \& Piddington 1938) 
might be attributed to 
cosmic ray EAS, although at the time both the energies and fluxes of 
such events were only crudely known. Later confirmation of the
meteoric origin of higher altitude ionization trails may have
contributed to 
a lack of follow-up on this prescient suggestion for the source
of these low-altitude sporadics. The idea was re-examined
in the early 1960's (Suga 1962) and a proposed experiment
was described by Matano et al. (1968); however,
no results from this test have appeared in the literature to date.

EAS ionization trails are now commonly detected by their 
air fluorescence
emission at near-UV wavelengths (cf. Baltrusaitas et al. 1985). 
Future large EAS detector arrays such as the Auger project 
(Gu\'{e}rard et al. 1998) and the proposed space
mission OWL/AirWatch\footnote{Recently, the AirWatch program has
evolved into a mission under consideration for the International
Space Station, and has been renamed Extreme Universe Space
Observatory (EUSO).}
 (Scarsi et al. 1999; Krizmanic et al. 1999)
have made fluorescence detection of EAS a centerpiece of
their approach, since it can provide information such as the
position of the shower maximum and the total shower energy which
can be difficult to pin down with particle detectors alone. 

One of the most compelling reasons
to extend the sensitivity of EAS detectors in this energy regime is
the possibility that neutrinos of energy $\geq 10^{19}$ eV may be
an important component of the primary particles (Capelle et al. 1999),
contributing to the fluxes at these energies either directly
or indirectly. Neutrinos and other highly penetrating particles
will induce air showers that develop much deeper in the atmosphere
than hadron-initiated showers. A unique signature of their presence
would be the detection of highly inclined (near-horizontal) showers
for which the point of maximum development is many radiation lengths
deeper than is possible for a hadronic or electromagnetic shower.
Thus any technique which (like the air fluorescence approach) has
a particular sensitivity to showers of this type is of interest
in addressing this problem.

To our knowledge no one
has reported any studies the ionization trail of EAS
using radar echo techniques, although as we will show here, the signals
should be clearly detectable using standard radar methods.
One reason for this may be the transient nature of the
expected echoes; short duration transients are typically
rejected by radar systems designed to detect or track targets
where the reflectivity does not rapidly decay. 

\paragraph{Ionization columns: meteors vs. EAS.}

Although there are many significant differences between the 
ionization caused by meteors and that of EAS, we can use the
basic formalism developed for meteor study as a starting-point
for analysis of EAS ionization. 
Meteor ionization trails are commonly parameterized in terms of their
ionization line density $\alpha$ (electrons m$^{-1}$), a measure of the 
total ionization content divided by the length of the meteor track.
Typical radar-detected meteors occur at heights of 80--120 km, and
have line densities of $\alpha = 10^{13}$ to $10^{16}$ m$^{-1}$.
At the lowest detectable line densities, the incident meteor has
a mass of $\sim 1~\mu$g, with a radius less than 100 $\mu$m. 
At typical velocities of $\geq 30$ km s$^{-1}$,
the implied kinetic energy of these meteor grains is 0.05 Joules or more,
much of which goes into ionization of the air along it path.

A cosmic ray proton of energy $10^{18}$ eV also has a
kinetic energy of order 0.1 J, and much of this energy also ultimately
ends up in the form of ionization and excitation of atoms of the air
along the path of the shower of charged particles that results from
the proton's collision with nuclear hadrons. The primary differences
between the meteor track and cosmic-ray-induced 
EAS are in the way the ionization
column forms, and in the resulting ionization density profile.

For the meteor, ablation of material from its surface 
yields atoms with kinetic energies of $10^2-10^3$ eV,
which ionize air molecules by direct collision with a
mean free path of several cm. This produces an ionization column with
an approximately uniform distribution of radial density, 
and an initial radius of 2-10 m.
The density then evolves with time due to
diffusion, convective processes, bulk motions of the air,  
and the Lorentz forces of the ambient electric and geomagnetic fields.
Electron attachment and recombination eventually complete the
process of dissipating the ionization column.
At radar frequencies in the HF to lower VHF range (10-100 MHz), echoes from
typical meteors may be detectable for several seconds after the
meteor is gone.

The ionization in an EAS, in contrast, is not produced by a single
body, but rather by the collective effects of the disk highly energetic
particles (mostly electrons and positrons) that make up the body of the
shower. Because the lateral distribution of these particles spreads
out as the shower progresses, the ionization column 
is formed with a different initial
distribution than that of a meteor, reflecting the evolution of the
cross-sectional charged particle density. In addition, since the shower
propagates essentially at the speed of light, it appears almost
instantaneously compared to even the fastest meteors at $\sim 100$ km s$^{-1}$.

An important measure of the transverse charged particle
distribution in an EAS is the Moliere radius
$r_m$ within which of order 90\% of the charged particles can be found.
For air at sea level, $r_m \simeq 70$ m, but it is important to note
that within $r_m$ the radial distribution is a power law, and
most showers retain a tight core of particles of diameter several m or less
which may look much like the column initially produced by a meteor. 
In fact, of order 10\% of the total number of electrons in an EAS
are likely to be found within a radius of $\simeq r_m/20$, comprising
much less than 1\% of the area of the Moliere disk.

\paragraph{Lightning ionization columns.}
Lightning discharges also produce intensely ionized regions
which have also been the subject of many radar studies since the early
1950's (Ligda 1950, 1956; Miles 1952). Ionization charge densities
produced by lightning are many orders of magnitude higher than
those produced by either EAS or small meteors, and the plasma channels
produced by lightning are also of much smaller diameter than
either meteor or EAS ionization columns. However they do
occasionally occur in clustered networks with
transverse scales comparable to those of EAS and meteors, and at
altitudes that overlap the altitudes of EAS maxima. Thus they can provide
examples of radar targets with features that are in some ways
relevant to our discussion.

\vspace{0.5cm}

In the following section we present the concept of the radar
cross section and outline specific cases relevant to
EAS ionization columns. Section 3 develops a semi-analytical method for 
determining the ionization densities for a given air shower energy
and altitude, and discusses some issues related to the evolution of
the electron density. Section 4 then makes estimates of the radar echo power
based on the formalism developed in the
previous two sections. Section 5 concludes the paper with a
discussion of some applications to specific experimental conditions.

\section{Radar detection of ionization columns}

For radar detection of the columns that result from meteors,
EAS events, or lightning, there are two regimes to consider, depending on the
plasma frequency $\nu_p$ of the ionized region:
\begin{equation}
\label{plasfreq-eq} 
\nu_p = \sqrt{(n_e e^2/\pi m_e)} \simeq 8.98 \times 10^3 \sqrt{n_e}~{\rm Hz}
\end{equation}
where $n_e$ is the electron density in cm$^{-3}$. 
These two regimes are known in
as the under- and over-dense regimes,
respectively, and traditionally (in the meteor radar literature)
are divided at the line density of
$\alpha \simeq 10^{14}$ m$^{-1}$. However, because of the
significant differences in the radial distribution of the 
electrons in EAS ionization compared to that of meteors, we distinguish them
only on the basis of the ratio of radar frequency to plasma frequency,
a distinction that is also common to radar lightning measurements.
Thus we will consider the over-dense portion of an ionization
column to be that region where the electron density is high enough that
the plasma frequency exceeds the radar frequency, and the radar
cannot penetrate it and is reflected. Underdense
columns are those which have electron densities such
that the local plasma frequency is below the frequency of the incoming
radar, which can therefore penetrate the ionized region.
\footnote{We stress here that underdense ionization columns may still 
produce a large radar return, for although the incident
radio waves penetrate the column, the scattering may still
be largely coherent, yielding a return power that scales
quadratically with the electron density. This feature is
of particular importance in determining radar's ability
to provide estimates of the shower electron density and thus ultimately
the energy of the primary particle.}

Radar targets are most commonly described in terms of their effective
radar cross section (RCS), a measure of the equivalent physical area of
an ideal scattering surface. Here we will develop concepts useful
to understanding the RCS 
characteristic to ionization columns, to prepare for later
estimation of the expected radar return power from EAS-induced ionization.

In the following subsections we introduce general formulas for the RCS for
both over- and under-dense cases. 
An additional distinction is also
useful in each case: the long wavelength, or {\em Rayleigh} regime
involves targets with characteristic dimensions smaller than the
radar wavelength used; and the short wavelength, or {\em optical} 
regime involves cases where the characteristic target size is much
larger than the radar wavelength.

\subsection{Overdense ionization columns}

When the electron density $n_e(r_c)$ is high enough to produce a surface where
the plasma frequency exceeds the frequency of the
incident radiation at some critical radius $r_c$ from the track, the resulting
index of refraction  becomes imaginary, and total external reflection 
of the radiation occurs. Under these conditions
the surface at $r_c$  can be treated to first order as a 
metal cylinder, and the RCS
is accordingly greatly enhanced. 

\subsubsection{Overdense columns: Rayleigh regime}

When $r_c \ll \lambda$, the cylindrical region
of reflecting plasma can be approximated as
a thin wire. As we will show later when we develop an estimate of the
ionization density for EAS, this is the case that will be most relevant to
low-frequency radar echoes from EAS. In this case the maximum of the RCS,
which occurs for normal incidence, can be
approximated as (Crispin \& Maffett 1965):
\begin{equation}
\label{thinwire-max-eq}
\sigma_{b,max}^{od} ~\simeq~ {\pi L^2 \cos^4 \phi \over
  \left ( {\pi \over 2 }\right )^2 ~+~ 
\left ( \ln [\lambda / (1.78 \pi r_c)] \right )^2 }
\end{equation}
where $\phi$ the angle of linear polarization with respect to the axis of
the wire. The behavior of $\sigma_b$ at angles other
than normal incidence is given by:
\begin{equation}
\label{thinwire-eq}
\sigma_{b}^{od}(\theta) ~\simeq~ {\lambda^2 \tan^2\theta \cos^4 \phi \over
16 \pi  \left [ \left ( {\pi \over 2 }\right )^2 ~+~ 
\left ( \ln [\lambda / (1.78 \pi r_c \sin\theta)] \right )^2 \right ]}
\end{equation}
where $\theta$ is the angle as measured from the axis of the wire
(or EAS in our case). This approximation is valid in the range
$60^{\circ} \leq \theta \leq 120^{\circ}$, but underestimates the
cross section at angles closer to the axis of the wire, due to
what are known as {\em traveling wave} effects, which produce 
large cross sections near the axis, in some cases comparable to that
at normal incidence (Medgyeshi-Mitschang \& Putnam 1985), and significantly
broader in angular range. However, the treatment of such effects is
very complex and beyond our scope at present.

\subsubsection{Overdense columns: Optical regime}

When $r_c \gg \lambda$, the reflection from the critical surface
becomes almost entirely specular, and for normal incidence we have
(Kraus 1988)
\begin{equation}
\sigma_{b,max} ~=~ {2\pi r_c L^2 \over \lambda} 
\end{equation}
where $\sigma_{b,max}$ denotes the maximum value of the RCS, and
$L$ is the length of the cylindrical reflecting surface.
Radar cross sections for metal cylinders have been studied in
detail (cf. Medgyeshi-Mitschang \& Putnam 1985) and are
complicated functions of angle, with resonances and nulls that
depend strongly on the length and other factors. 
As we shall establish below, 
EAS ionization columns will generally not fall into the
optical regime for the overdense case, except at very
high EAS energies and very low radar frequencies.
For this reason we do not treat these effects further here.

\subsection{Underdense ionization columns}

For radar frequencies well above the plasma frequency $\nu_p$, 
the radiation penetrates the
entire ionization column, and
the electrons within the ionization column scatter independently according
to the Thomson cross section
\begin{equation} 
\sigma_{T} = {8\pi \over 3} \left ( {e^2 \over m_e c^2} \right )^2
~=~ 6.65 \times 10^{-29} {\rm m^2}~.
\end{equation}
The total effective radar backscatter cross section $\sigma_b$ will then depend
on the individual phase factors of each of the scattering electrons. 

\subsubsection{Underdense columns: Rayleigh regime}

For $r_m \ll \lambda$, and
angles that are nearly perpendicular to the track, 
all of the electrons scatter coherently over a longitudinal region of the track 
$L_F= \sqrt{\lambda R/2}$ where $R$ is the perpendicular distance 
to the track and
$\lambda$ the radar wavelength (Hanbury-Brown \& Lovell 1962). 
This region of the track is known as the
first Fresnel zone of the track, and the total radar backscatter cross
section then becomes 
\begin{equation}
\label{sig-th}
\sigma_b = N_e^2\sigma_{T} ,
\end{equation} 
where $N_e = \alpha L_F = \alpha \sqrt{\lambda R/2}$ is the number of electrons
within a single Fresnel zone along the track. The radar cross
section depends of the square of the electron density because
of the assumption of full coherence. 

Except for radar frequencies $f \ll 10$ MHz, most radar ranging of
EAS ionization columns is not in the Rayleigh regime.
However, the concept of the Fresnel zone, which establishes a
characteristic length scale over the longitudinal extent of the track,
will still be useful in what follows.

\subsubsection{Underdense columns: Optical regime}

For tracks where the radar frequency $f > \nu_p$, but 
$r_m \geq \lambda/4$, the assumption of
coherent scattering is no longer satisfied, and the individual
phase factors of the electrons must be included. 
For this case the
effective RCS can be written (cf. Wehner 1987;
also see appendix):
\begin{equation}
\label{phasor-eq}
\sigma_b^{ud} ({\bf k}) ~=~ \left | 
~\int n_e({\bf r})\sqrt{\sigma_{T} }e^{2i{\bf k \cdot r}} 
d^3{\bf r} \right |^2
\end{equation}
where ${\bf k}$ is the wave vector of the incident field ($k=2\pi/\lambda$), 
${\bf r}$ is the vector distance to the volume element at which
the scattering takes place. Here we are neglecting refractive 
effects of the ionization column on the incident radiation, and
assuming that the incident and backscattered waves satisfy 
${\bf k}\cdot {\bf r} \simeq |kr|$.

Note that the argument of the exponential here
includes an extra factor of two to represent the two-way phase.
This is due to the fact that as the scattered radiation returns on the
same path as the incident radiation, it picks up an
additional phase factor equal to that of the incoming radiation.
The term $n_e({\bf r})\sqrt{~\sigma_{T} }~d^3r$ represents the 
differential contribution to the scattered electric field
of a volume element of electrons which scatter coherently.

If we now write ${\bf q} = 2{\bf k}$ then equation \ref{phasor-eq}
becomes
\begin{equation}
\label{fft-eq}
\sigma_b^{ud} ({\bf q} ) ~=~ \sigma_{T} \left | 
~\int  n_e({\bf r})e^{i {\bf q \cdot r}} d^3{\bf r}
\right |^2  \\
\end{equation}

Equation \ref{fft-eq} thus reduces the problem of estimating the effective
cross section in the underdense case to that of calculating the
Fourier transform (and the resulting power spectrum)
of the electron number density distribution. In the appendix we 
describe the geometry
more explicitly and show how this relation arises.

A broader implication of
this result can be stated as follows:
{\em A measurement of the 
complex amplitude of the radar echo from an EAS in the underdense
case is proportional to a measurement of one Fourier component of the
shower ionization profile.} This will be true even in the
case of a bistatic radar (receiver location not coincident
with transmitter) although there will be an additional coefficient
to account for the bistatic cross section and its angular behavior.

\section{EAS ionization densities}

In the previous section we introduced the mechanisms for radar reflection 
from cylindrical plasma columns in two regimes of ionization
density. Here we estimate the spatial distribution of the
expected ionization from a high energy
cosmic-ray air shower, which will then allow us to assess their effective
radar cross section.

\subsection{Longitudinal ionization}

There are many years of development of the theory of EAS production. 
The most accurate treatments of the evolution of the electron density
in the shower require numerical simulations, but there are a number of
parameterizations available that yield results accurate enough for
our needs. Here we use the analytical model due originally to
Kamata \& Nishimura (1958), and Greisen (1965), often referred to
as the NKG approximation.
In the NKG model the longitudinal (along-track)
development of the shower at a depth $d$ is parameterized by 
its {\em age} $s$:
\begin{equation}
s(d) ~=~ { 3 d/X_0 \over d/X_0 + 2\ln(E/E_{crit}) }
\end{equation}
where $E_{crit}=86$ MeV for electrons in air, and $X_0=36.7$ gm cm$^{-2}$ is
the electron radiation length in air. The evolution of the total number of 
charged particles (virtually all electrons and positrons) with depth 
is then approximated by
\begin{equation}
\label{ne-eq}
N_e ~=~ {0.31 \exp [(d/X_0)(1-1.5\ln s )] \over 
\sqrt{\ln (E/E_{crit})}  }~.
\end{equation}
Although this approach
does not account for any of the large fluctuations that are possible
in high energy air showers, it describes the average behavior reasonably
well.

\placefigure{ne}
\begin{figure}[t]
\plotone{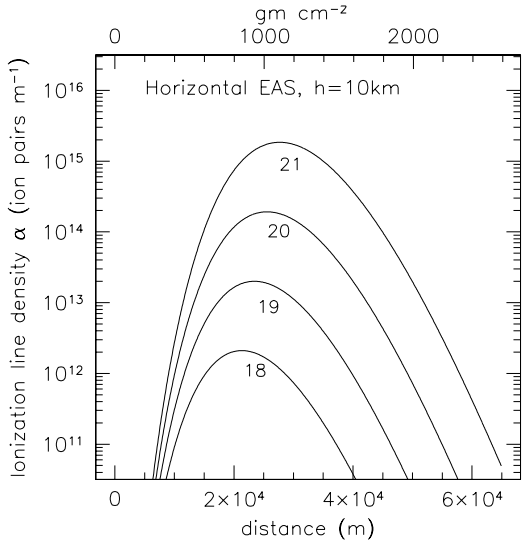}
\caption{Electron ionization line density for 5 showers of
energies in the $10^{18}$ to $10^{21}$ eV range. Such line densities are
quite similar to those of radio meteors.\label{ne-fig} }
\end{figure}

\subsection{Transverse ionization density}

The transverse charged particle
densities are described in a similar fashion, also parameterized
by the age of the shower and the Moliere radius $r_m$ (Bourdeau 1980):
\begin{equation}
\label{rhoe-eq}
\xi_e ~=~ K_N~\left ( {r \over r_{m}s_m } \right )^{s-2}
\left ( 1 + {r \over r_{m}s_m } \right )^{s-4.5}
\end{equation}
where
\begin{equation}
K_N ~=~ { N_e \over 2\pi r_{m}^2 s_m^2} ~
{\Gamma(4.5-s) \over \Gamma(s)\Gamma(4.5-2s) }
\end{equation}
and $\Gamma$ is the gamma function, and $s_m = 0.78-0.21s$.
The Moliere radius for air is given by 
\begin{equation}
r_{m} ~=~ 2.12\times 10^5~ {X_0 \over E_{crit}~ \rho}
~=~ 70 \left ( \rho_0 \over \rho \right )~{\rm m}
\end{equation}
where $\rho$ is the density of the air at the altitude under consideration,
and $\rho_0$ is the sea level density.
The calculated density $\xi_e$ is in units of charged particles per unit area,
passing through a plane transverse to the shower axis at the given
depth. Since the shower thickness 
is typically of order several meters or less over its length,
$\xi_e$ can be approximately equated with the number density in a slab 
transverse to the shower. 


To convert these particle number densities to the resulting ionization
density, we use the average value for electron
energy loss given by $E_{crit}/X_0 = 2.343$ MeV gm$^{-1}$ cm$^{-2}$,
divided by the mean energy per ion-pair for air $E_{ion} = 33.8$ eV
(Segre 1977) which accounts for the inefficiency of ion-pair
production. This procedure is similar to that
used by the Fly's Eye (Baltrusaitas et al. 1985) in estimating
shower parameters based on fluorescence yield. 

We note here that the NKG approach and subsequent modifications
to it do not accurately account for the electron distribution
at large radii from the core, and will in general underestimate
the electron density beyond the Moliere radius. However, 
since the ionization density at large radii is low compared to
the core, there is very little coherent contribution from these electrons,
in spite of their large integrated number.
Our results remain basically unchanged by this effect.

\subsection{Numerical results}

Fig. \ref{ne-fig} 
shows curves of calculated ionization line densities for several
EAS of different primary energies over the range of $10^{18}-10^{21}$ eV.
The showers are assumed to be propagating
horizontally at an altitude of $10$ km, and the shower parameters 
are corrected for the lower air density at this altitude. 
The bottom axis shows the 
along-track distance corresponding to the depth shown along the top axis.
As noted above, the line densities at these energies correspond to
typical line densities of radio meteors, which are detected at 
heights of 80-120 km. However, the lateral distributions will be
quite different as noted above.
We note also that the showers are physically very long,
with the region of shower maximum extending over many km of physical
distance for most showers at these energies.

\placefigure{cumne}
\begin{figure}[t]
\plotone{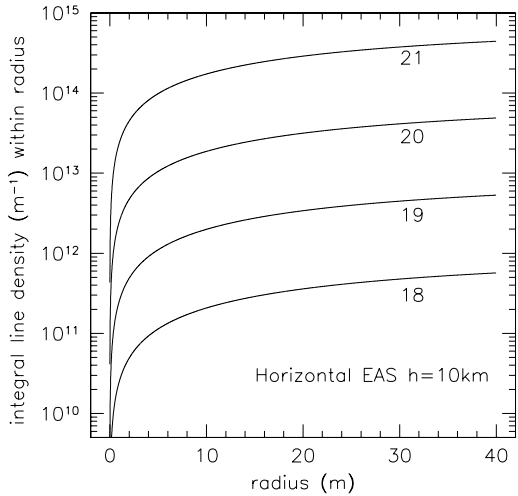}
\caption{The effective line density of an EAS near its shower
maximum as a function of the radius within which the electron
density is integrated. It is evident that the effective
line densities within radii of 10 m or so are about 10\% of the
total line densities shown in Fig. 1, and are still substantial. 
\label{cumne-fig}}
\end{figure}

In Fig. \ref{cumne-fig}  
we show the effective line density as a function of the
radius within which the electron density is integrated, for the
inner 40 m radius for the same set of showers shown in Fig. 1.
Here it is evident that the effective line densities
of the inner $\sim 10$ m radius core of the showers are about
a factor of 10 lower than those which include all the
electrons produced in the shower. However, this area corresponds to
only 0.25\% of the area of the Moliere disk ($r_m\simeq 200$ m at 10 km), 
It is thus evident that the cores
of the showers are highly concentrated in ionization density with respect
to the bulk of the shower disk area.

\begin{figure}[t]
\plotone{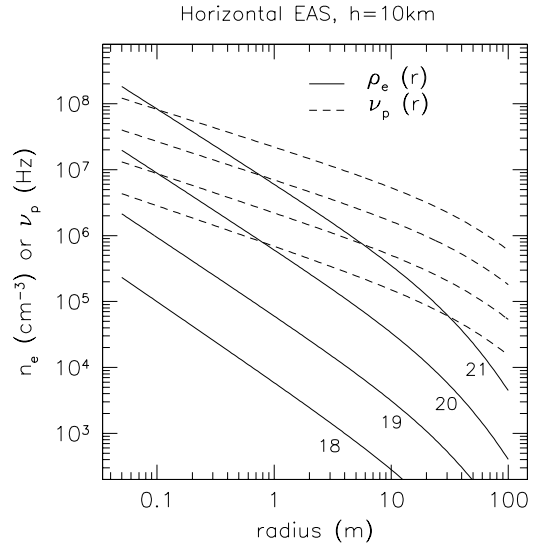}
\caption{Radial dependence of the ionization density for the same
showers presented in the previous figure. Also shown is the
radial dependence of the plasma frequency for each case.\label{rhoe-fig}}
\end{figure}

In Fig. \ref{rhoe-fig}, we show the lateral ionization density distributions 
near shower maximum for
the same four EAS presented in Fig. \ref{ne-fig} (solid lines). 
Included also are curves of the
effective plasma frequency corresponding to the density at each radial
distance (dashed lines).
Given the uncertainty in the accuracy of the
analytical model for EAS development at very small core radii (cf. 
$r<20$ cm), the implied highest radar frequencies that will undergo total
reflection are in the range of 10-50 MHz, with a strong dependence on
the primary energy. 

\placefigure{rhoe}

\begin{figure}[t]
\plotone{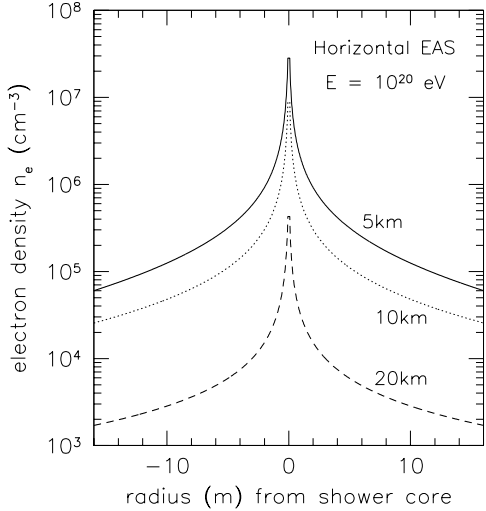}
\caption{Radial profile of showers shown for the inner 40 m radii,
plotted on a linear distance scale, for different shower
altitudes, showing the effects of altitude for a set of
horizontal showers.\label{rhoevsalt-fig}}
\end{figure}

It is evident from Figures \ref{cumne-fig} and 
\ref{rhoe-fig} that the cores of the showers
are likely to be most important in determining the radar
response, since the electron density is much higher near the
core. In Fig. \ref{rhoevsalt-fig}, we plot the core electron density again 
near shower maximum for
a $10^{20}$ eV horizontal shower, at a variety of different
altitudes to display the effects of the different atmospheric
density. To first order the shape of the core is preserved with
altitude, but the overall electron density decreases at higher
altitudes because the production rate of ionization depends on
the density of the air.

\placefigure{rhoevsalt}

\subsection{Duration of the radar echo}

Before estimating the radar echo power, we first address the
important topic of the expected lifetime of the free electrons
in the ionization column. If EAS detection by radar is to be
proven practical, 
it is first necessary to first establish that the ionization
column can be treated as quasi-static over the timescale it might
take to interrogate the shower with a radar pulse, either as a
standalone system, or
if a separate air shower detector provided the trigger. 

As an example of the issue with respect to triggered operation, 
consider a large air shower array with radius
30 km. For typical electronics used in such systems the trigger forms within
several tens of $\mu$s. Round trip 
transit time to an EAS within $\sim 30$ km requires another $200~\mu$s. 
Thus in this case we require EAS electrons to remain mostly free
for of order 250 $\mu$s or longer, if an external trigger is to be
used to allow interrogation of a detected shower. In general,
if the free electron lifetime is $\tau_e$, and the trigger formation
and pulse initiation time is $\tau_p$, then the maximum range for 
a triggered system is
\begin{equation}
R_{tr} ~\le~ c ~{\tau_e - \tau_p \over 2}~.
\end{equation}

If a radar detection system is operated in a standalone mode, the
free electron lifetime will determine how often the detection
volume must be pulsed in order to have a reasonable efficiency at
detecting the EAS. In the limit where the free electron lifetime 
becomes so short that it is comparable to the pulse duration, 
then pulsed radar detection will become impractical. Other
methods which use continuous-wave (CW) techniques may still
apply in this case; these will be discussed briefly in a later
section.

There are a number of effects that cause the electron density to evolve
with time:
\begin{enumerate}
\item Diffusion of the electrons through the ambient air, lowering their
density and thus their net radar cross section;
\item Electronic recombination with ions;
\item Attachment of electrons to neutral molecules, or dissociative 
attachment to atoms;
\item Collisional detachment of previously attached electrons, which
frees electrons that have been attached to atoms;
\end{enumerate}

\subsubsection{Diffusion effects}

Although the conditions of the atmosphere at the 80-120 km heights that
meteor echoes are observed at are quite different from those at
EAS heights, it is useful to preface our analysis with a discussion of
the effects that determine the time scale for meteor echoes.

The duration of meteor radar echoes has been modelled and studied
experimentally in detail for many decades 
(cf. Hanbury Brown \& Lovell 1957; Kaiser 1968;
Kaiser et al. 1969; Jones \& Jones 1990; Jones 1991). The power
of a meteor radar echo is found in the underdense regime to
decay exponentially with a time constant 
$\tau_m = \lambda^2 /(32\pi^2 D_{i})$
where $D_{i}$ is the ambipolar (or ion neutral) diffusion coefficient. 
At the altitudes that meteors are observed with radar, 
$D_{i} \simeq 1-10$ m$^2$ s$^{-1}$, and the typical decay
times for underdense trails 
are thus several tens of ms for frequencies in the VHF
regime. 

To estimate diffusion effects at the lower altitudes
of EAS ionization columns, we note that $D_{i} \propto T\nu_i^{-1}$
where $T$ is the kinetic temperature and $\nu_i$ the collision
frequency (cf. Buonsanto et al. 1997). Thus at 10 km altitude, 
the diffusion coefficient is much smaller, 
$D_{i} \sim 5$ cm$^2$ s$^{-1}$, due
mainly to the much higher collision frequency at lower altitudes.
The implied time constant, assuming diffusion progresses in the
same manner for EAS as for meteor ionization columns, is
$\tau_{eas} \simeq 60 $ s for $\lambda = 3$ m.

S-band radar echoes from lightning at altitudes of $\sim 10$ km
measured by Williams et al. (1984) were frequently found to
persist for several hundred milliseconds. For a wavelength of
11 cm, the diffusion time constant at this altitude is 240 ms, 
which suggests that diffusion
may be the dominant effect in the evolution
of lightning radar cross section, at least at microwave frequencies.
However, lightning ionization densities are estimated to be
initially $10^{13}-10^{14}$ cm$^{-3}$, with plasma frequencies
$10^{3}$ or more times higher than for EAS columns. Thus other
effects relevant to the conditions of this much denser plasma
may be important.

\subsubsection{Attachment \& Recombination}

Since diffusion is less effective at dissipating the ionization column
of an EAS than it is for a meteor, at least at VHF frequencies,
we expect that attachment and recombination
may play a more significant role. 
The recombination rate for 
atmospheric electrons is described by (cf. Thomas 1971):
\begin{equation}
\label{recomb-eq}
{ dn_e \over dt } ~=~ N_e^0 -\alpha_e n_e^2 -\beta_e n_e
\end{equation} 
where $N_e^0, ~n_e$ are the initial and evolving electron density, 
respectively, $\alpha_e$ is
the electron-ion recombination coefficient
(in units of cm$^{3}$ s$^{-1}$), $\beta_e$ and is the
electron attachment rate (s$^{-1}$). 
Here we have ignored 
terms in Thomas (1971) which relate to the presence of negative ions
in the ionosphere
since we are considering EAS within the troposphere
where such ions are not prevalent. 

The  
electronic recombination coefficient 
for hydrogenic atoms can be
approximated by (Seaton 1959) 
\begin{displaymath}
\alpha_e ~=~ 5.20 \times 10^{-14}~ Q^{1/2}
\end{displaymath}
\begin{displaymath}
~~~~~~~~~\times (~0.429+0.5 \ln Q + 0.469 ~Q^{-1/3}~) {\rm~cm^{3}~ s^{-1}}\\ 
\end{displaymath}
where $Q=1.58 \times 10^5/T_e$
and $T_e$ is the electron kinetic temperature, with $T_e \simeq 10^3-10^4$ K
typical for ionization in this case. 
For air in the troposphere we
thus expect $\alpha_e \sim 10^{-12}$ to $10^{-11}$ 
cm$^{3}$ s$^{-1}$, and
the implied recombination time for the highest electron densities
we have considered is several minutes or more. Thus it is highly
probable that the electrons will attach before recombining.
\footnote{We note that 
values for the total recombination coefficient of
$\alpha = 2 \times 10^{-7}$ cm$^{3}$ s$^{-1}$
are found in the D region of the
ionosphere (Thomas 1971).
Such a high value is likely to be due to the large variety of 
ion species present, which increases the total
recombination cross section per electron.}

Molecular oxygen is electronegative,
forming ions with free electrons primarily through dissociative
attachment, via the process
\begin{equation}
{\rm e^- + O_2 \rightarrow O^- + O }
\end{equation}
where the resulting oxygen ion and atom carry off the excess
kinetic energy of the electron, and the electron affinity of O$_2$ is
of order 0.5 eV. Molecular nitrogen, in contrast, does not form
attachments with electrons but can be more easily collisionally excited,
and may play a role in mediating detachment processes in air.

Several studies of electron attachment in 
air using electron swarm techniques
(see Gallagher et al. 1983 for a compilation) 
have indicated a much lower attachment
rate than that which is expected from using the attachment
coefficient for pure molecular oxygen reduced by the molar ratio in air
(Moruzzi \& Price 1974 \& references therein).
Measurements of the attachment coefficient are very sensitive to 
contamination by carbon dioxide, and CO$_2$ was not removed from the
samples in many early measurements. Although CO$_2$ can be 
present in significant concentrations 
in the near sea level atmosphere, its average 
concentration in the atmosphere as
a whole is $0.033$\%. Thus electron attachment rates for dry, high-altitude
air must be determined with CO$_2$ largely removed from typical
air samples.

Moruzzi \& Price (1974),
after carefully removing all CO$_2$ from their samples, confirmed
earlier measurements in being unable to
detect any attachment in pure dry air. This result was attributed to 
a competing process: 
rapid collisional detachment from N$_2$ interactions, yielding an
effective attachment rate that was less than 10\% of the 
ionization rate and thus too small too measure. Their
limit on the effective 
attachment coefficient $(\eta^*/N_m)$ for pure air is
\begin{equation}
(\eta^*/N_m) ~\leq~ 3 \times 10^{-20} ~{\rm cm^{2}}.
\end{equation}
where $N_m$ is the number density of the attaching molecular
species. The attachment rate $\beta_e$ is related to the 
effective attachment coefficient through
\begin{equation}
\beta_e = (\eta^*/N_m(h)) u_e N_m(h)
\end{equation}
where $u_e$ is the electron drift velocity. For a typical electron
drift velocity of $u_e = 2 \times 10^5$ cm$^{-1}$ s$^{-1}$,
the limit on $\eta$ gives $\beta_e \leq 5.4 \times 10^4$ s$^{-1}$,
implying that attachment dominates completely over recombination.

Figure \ref{attach-fig} shows the behavior of the electron density
for three different altitudes for a $10^{19}$ eV shower core region.
Here we have plotted the electron density time evolution for both
the case where the attachment rate is equal to the limit of
Moruzzi \& Price (1974), and one-tenth of the limit (dashed lines).
It is evident that the expected electron lifetimes are at least 
10-20 $\mu$s, increasing slowly at the higher altitudes.

\placefigure{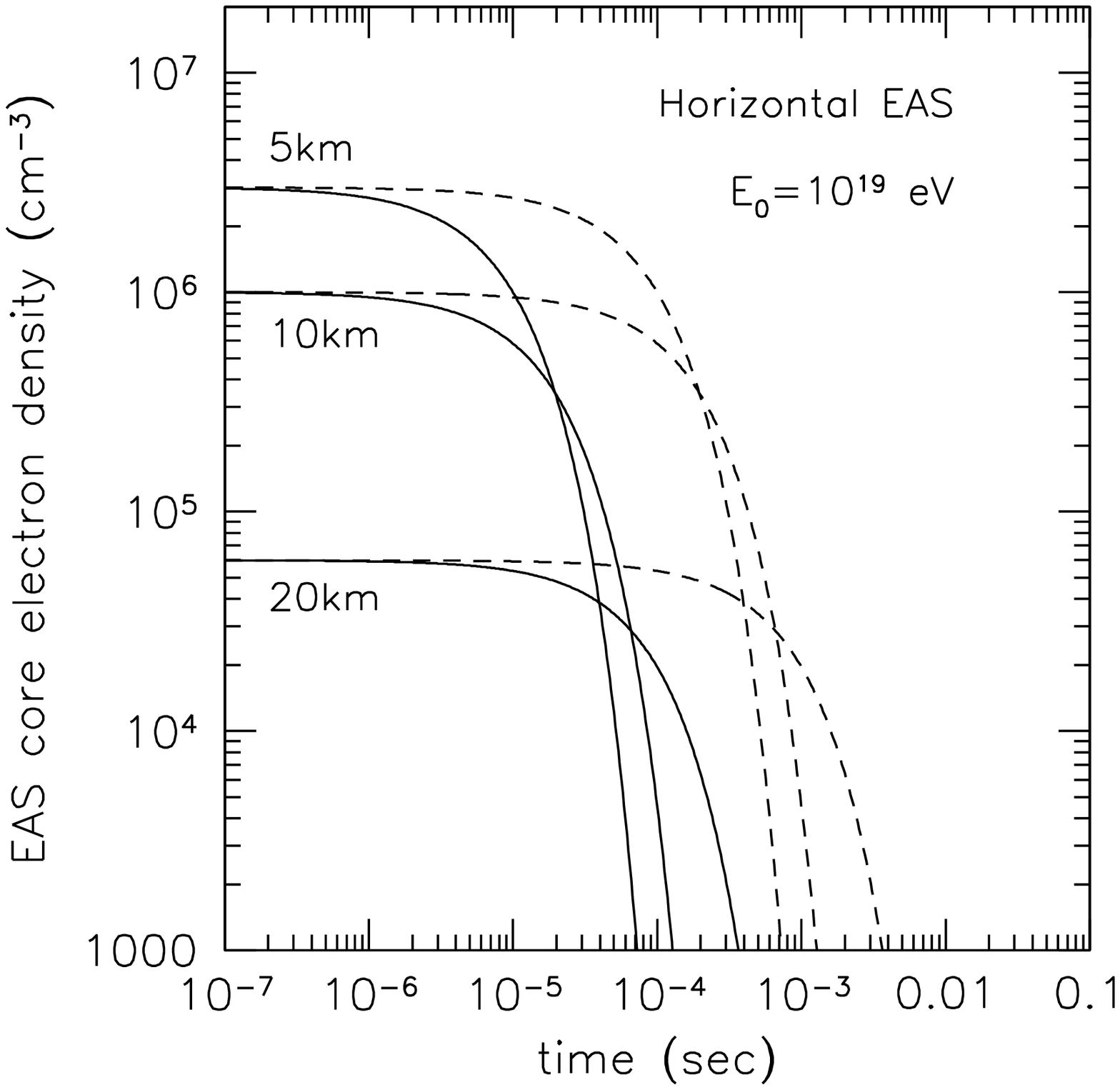}
\begin{figure}[t]
\plotone{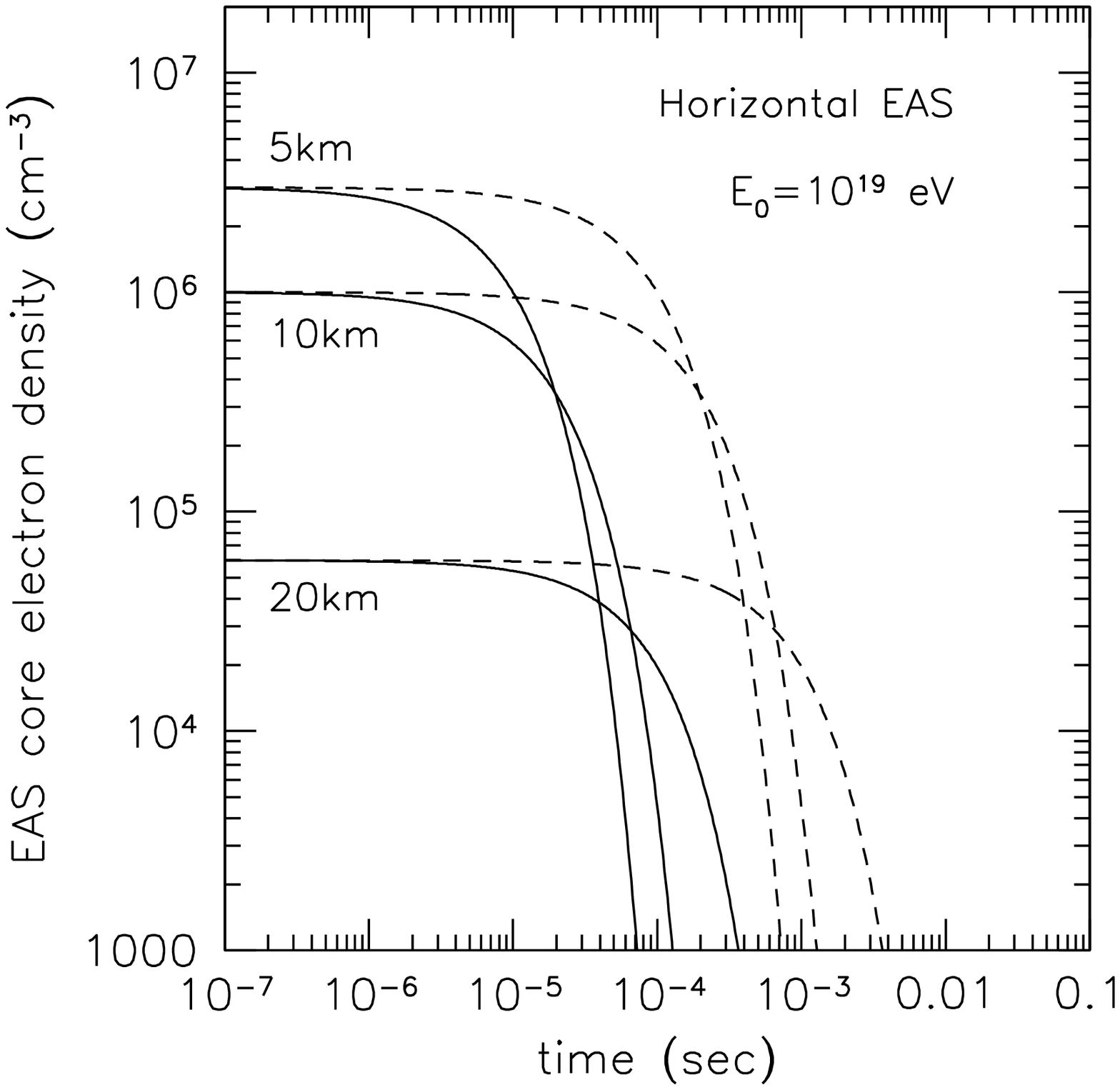}
\label{attach-fig}
\caption{The evolution of the electron number density
is shown for three $10^{19}$ eV EAS, at different altitudes.
The solid lines correspond to the evolution for the present
upper limit on the attachment rate, and the dashed lines
to one-tenth of the upper limit.\label{attach-fig}}
\end{figure}

Although this estimate of the electron attachment lifetime is
important from the point-of-view of evaluating whether there is
sufficient time for triggered-interrogation of an EAS with radar,
it is also important to establish an upper limit on the 
electron lifetime, both for purposes of evaluating the possibility of
multiple-pulse interrogation of a shower, and for 
considerations of radar clutter from echoes of multiple showers 
for a radar system with very large range. 

The lightning radar measurements noted above (Williams et al. 1984
\& references therein) often found radar echo persistence of several
hundred ms using repetitive radar pulses. 
However, the
plasmas produced by lightning are extremely overdense and hot relative
to those we are considering, and thermal detachment may play
a significant role. In addition, the large amount of water
vapor present in typical lightning radar measurements 
complicates any comparison with EAS in dry air. Thus comparison
of radar echoes from lightning with those from EAS ionization 
must be done with caution.

An rough upper limit for the free electron lifetime can be
obtained by considering the equilibrium level of free electron
content of the mesosphere in the region of maximum cosmic ray
ionization, peaking at about 12-14 km altitude. 
Here the dominant ionization is 
caused by low energy (several hundred GeV and above) cosmic ray air showers,
which produce of order 30 ion pairs per cm$^{3}$ per second
(Rossi 1966; Gregory \& Clay 1982), roughly 15 times the sea level
rate of ion production.

The DC conductivity of the atmosphere at 12 km is 
$\sigma_0 \simeq 4 \times 10^{-13}~\Omega^{-1} {\rm m^{-1}}$
(Dolezalek 1982). The cosmic ray ionization at this altitude
induces a tenuous plasma, with DC conductivity
\begin{equation}
\label{dc-sigma-eq}
\sigma_0 ~=~ {4\pi\epsilon_0} {\pi \nu_p^2 \over \nu_{en}}
\end{equation}
where $\epsilon_0$ is the dielectric permittivity constant,
and the electron-neutral molecule collision frequency $\nu_{en}$
can be written as
\begin{equation}
\nu_{en} ~\simeq~ 4 \times 10^{-10} \sqrt{T_e} ~N_m
\end{equation}
where $N_m$ is the molecule number density (cm$^{-3}$) at the altitude
considered. The electron temperature $T_e$ is uncertain but
likely to be in the range of 300-2000 K. At 12 km we thus expect
$\nu_{ei} \simeq 4 \times 10^{10}$ s$^{-1}$ for $T_e=300$ K.

Inverting equation~\ref{dc-sigma-eq}
for the plasma frequency, which in turn depends on the equilibrium
electron number density (equation 1), we find
\begin{equation}
n_e ~\simeq~ 35.5 ~\nu_{en} ~\sigma_0 ~{\rm cm^{-3}}.
\end{equation}
For the values determined above we have 
$n_e \leq 0.57$ cm$^{-3}$ which implies a mean lifetime of 
less than 20 ms. The actual electron lifetime will be
less than this since the detailed DC conductivity also includes
contributions from ionic terms which we have neglected here.
There will also be an altitude dependence on this upper
limit since it depends on the collision frequency.
In summary, we find that the available time for EAS interrogation
by radar is bounded at  10 km altitude approximately by
\begin{equation}
20~{\rm \mu s} ~\leq~ \tau_e ~\leq 20~{\rm ms}~.
\end{equation}
Thus, for the shortest values of $\tau_e$, triggered
interrogation of air showers appears to be impractical except
for ranges of a few km or less; for the largest possible
values, ranges of hundreds to even a thousand km or more are
not excluded for higher altitude showers. In the latter case,
no existing system can yet trigger on showers at these great 
distances, but planned space-based systems such as OWL/AirWatch
will have such capability.

The $\sim 20$ ms upper limit to the electron lifetime shows also that
radar clutter from echoes of multiple showers should be 
completely negligible. For example, even assuming an energy threshold
as low as $10^{16}$ eV could be attained out to 10 km, the
shower rate at this energy over this area is still only
several events per second. Radar systems also generally allow
range-gated triggers, and this can facilitate rejecting any
nearby clutter from low-energy showers within a few km of the
transmitter.

\section{Predicted radar return power from EAS}

Having established that the duration of the
EAS radar targets is likely to be sufficient for their detection, and
introduced the radar cross sections expected, we now turn to 
estimation of the return power of the echo.

Radar return power $P_r$ is described in terms of a model where
the radiation is emitted from an antenna with peak transmitted power $P_t$
and directivity gain $G = 4\pi\Omega_A^{-1}$,
where $\Omega_A$ is the solid angle of the main beam of the antenna.
The radiation is assumed to then scatter from objects in the
antenna beam and be re-radiated isotropically in the frame of the
scatterer, producing a $R^{-4}$ dependence in the returned power
as a function of range $R$. Deviations from isotropic scattering are 
thus absorbed into the effective radar backscatter cross-section $\sigma_b$,
which can be larger or smaller than the physical cross-section of
the object. In addition any real transmitting and
receiving system will have less than unity efficiency,
which we designate here as $\eta$.
The radar equation under these conditions is (cf. Skolnik 1990)
\begin{equation}
\label{rad-eq}
{P_r \over P_t } ~=~ \sigma_b \eta~{G^2 \lambda^2 \over (4\pi)^3 R^4}~.
\end{equation}
Here we are assuming that the transmitting and receiving antennas are
identical, and we are neglecting for the moment any polarization
effects or losses in the medium.
\footnote{This equation is also derived strictly under conditions
where the received radiation is in the far-field, that is, where
$R> 2D^2/\lambda$ where $D$ is the largest projected dimension of the
scattering target. In our case, this is not generally satisfied, since
the length of the ionization columns can be tens of km. However,
we have already accounted for these Fresnel zone effects by limiting our
analysis to the first Fresnel zone as noted above.}

Given equation \ref{rad-eq}, the problem of determining the detectability
of EAS-initiated ionization columns reduces to that of determining the
effective RCS $\sigma_b$ for a given choice of operating
radar frequency, and the noise power of the
specific radar in use. The noise power is given by
$P_N = kT_{sys} \Delta f$, 
where $T_{sys}$ is the system noise temperature,
$k$ is Boltzmann's constant, and $\Delta f$ the effective
receiving bandwidth, assumed here to be matched to the transmitting
bandwidth. 

\paragraph{Pulse compression radar.}
Almost all modern pulsed radar systems now use what is known
as {\em pulse compression}, a method which allows the receiver bandwidth,
and thus the noise power, to be minimized (cf. Wehner 1987). 
Pulse compression is typically
implemented by effectively dispersing a band-limited pulse
with an initial bandwidth $\Delta f_0 = (\Delta t_0)^{-1}$
through a filter, transforming it into
a frequency chirp which spans the original bandwidth, but now has 
a duration $\Delta t \gg \Delta t_0$. The receiver then uses an inverse filter
to de-disperse the received pulse. The effective bandwidth is thus
$\Delta f = (\Delta t)^{-1}$, but the range resolution 
$\Delta R \propto \Delta t_0$ of the
full bandwidth is recovered.

\paragraph{CW radar.} Continuous wave radar systems which transmit
at 100\% duty cycle, can derive range information by using
frequency- or phase-encoded waveforms to uniquely tag the 
detected echoes to a particular time segment of the transmitted
waveform. Processing of the received signal typically
involves mixing with a reference signal, which transforms 
different ranges to different frequencies, followed by
detection in a filterbank. The effective bandwidth is then
determined by the bandwidth of the frequency channels used.

For simplicity we will assume a pulsed radar system here and in
what follows.

Combining the noise power equation above with equation \ref{rad-eq}, the
signal-to-noise ratio (SNR) of the received power is
\begin{equation}
\label{sn0-eq}
{P_r \over P_N } ~=~ {\sigma_b P_t \eta} ~{G^2 \lambda^2 \over (4\pi)^3 R^4} 
~{1 \over k T_{sys}\Delta f } ~.
\end{equation}

Evaluating equation \ref{sn0-eq} for a nominal choice of
parameters gives the SNR per received radar pulse
per square meter of RCS:

\begin{displaymath}
\label{sn-eq}
{S \over N } ~=~ 3.3 \left ( {\sigma_b \over 1 ~{\rm m^2}} \right )
\left ( {P_t \over 1 ~{\rm kW}} \right )
\left ( {\eta \over 0.1} \right )
\left ( {G \over 10} \right )^2
\end{displaymath}
\begin{equation}
\times \left ( {\lambda \over 3 ~{\rm m}} \right )^2
\left ( {R \over 10^4 ~{\rm m}} \right )^{-4}
\left ( {T_{sys} \over 10^3 ~{\rm K}} \right )^{-1}
\left ( {\Delta t \over 10~ \mu {\rm s}} \right ).
\end{equation}

We note that the reference
values chosen here represent a modest radar system;
in particular the peak power of 1 kW is easily attained by current standards, 
and the directivity $G\simeq 10$ (beam size $\sim 1$ sr)
represents a relatively low-gain antenna. 
The system temperature $T_{sys}=1000$ K
is realistic for 100 MHz ($\lambda = 3$ m) however, 
since the brightness temperature of
the sky is quite high at these frequencies. 

Radar systems also
routinely use repetitive pulsing to increase SNR, which then grows
roughly as $N_p^{1/2}$ where $N_p$ is the number of pulses that are averaged.
In the case of an EAS, the number of pulse repetitions is limited
by the free electron lifetime, with large uncertainty. For
the shortest possible electron lifetimes, it is unlikely that
repetitive pulsing can provide much improvement in SNR,
but for lifetimes of order 1 ms, repetitive pulsing could
provide substantial improvements in SNR.

We now consider in detail the expected RCS for
EAS ionization columns in the various density regimes that were
introduced above.

\subsection{Low-frequency/overdense case}

\placefigure{thinwire}
\begin{figure}[t]
\plotone{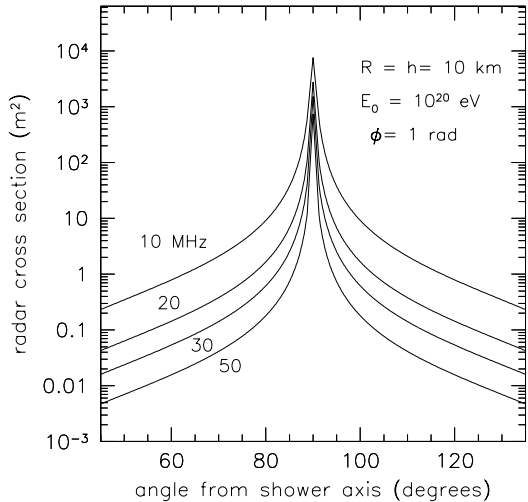}
\caption{Radar cross section over a range of angles centered
on normal incidence (at $90^{\circ}$) for an air shower with
$E_0=10^{20}$ eV, at a distance of 10 km, at radar wavelengths
30, 15, and 10 m. Results are 
based on the thin wire approximation. The polarization of the
incident wave has been taken to be 1 radian with respect
to the shower axis; thus these RCS values should apply
on average to random relative orientations.\label{thinwire-fig}}
\end{figure}

\placefigure{phasefac}
\begin{figure}[t]
\plotone{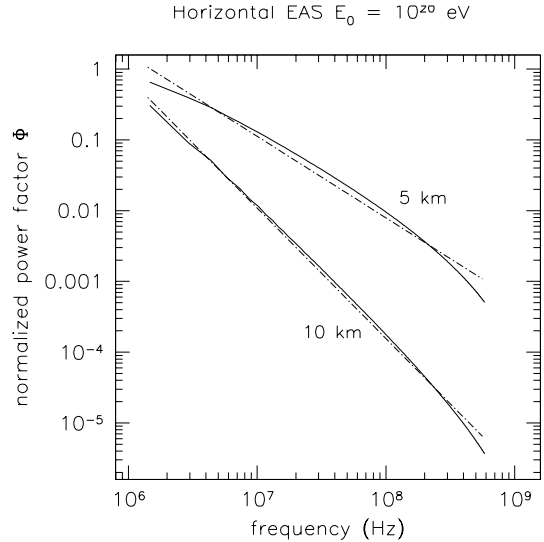}
\caption{Normalized power loss factor due to loss of coherence
for two horizontal E = $10^{20}$ eV showers at heights of 5 and 10 km.
\label{phasefac-fig}}
\end{figure}

At radar frequencies below the plasma frequency of the ionized core of the EAS,
the RCS is comparable to that of a metal cylinder,
as noted above, with a radius of $r_c$ and
a length equal to that of the Fresnel zone. 
Based on the analytical shower model used above and the results in
Fig. \ref{cumne-fig}, the critical radius at a
given wavelength and EAS cascade energy $E_0$ 
can be empirically approximated by
\begin{equation}
{r_c \over \lambda} ~=~ {1 \over 30}~
\left ( {f \over 10 ~{\rm MHz}} \right )^{-0.75}
\left ( {E \over 10^{20}{~\rm eV}} \right )^{0.85}~{\rm ~m}.
\end{equation}
It is evident from this result that for frequencies above
10 MHz, and for all but the highest EAS energies considered here,
$r_c \ll \lambda$ and we can use the thin wire approximation discussed above.

In Fig. \ref{thinwire-fig} we show results for radar cross
sections of the first Fresnel zone for a shower of $E_0 = 10^{20}$ eV,
for three different frequencies, for a range of angles centered on
normal incidence, and a polarization angle of 1 rad with respect to the
shower axis.
The cross section at normal incidence is
greatly enhanced due to the specular reflection at that angle.
We can empirically express the maximum cross section for this
case as

\begin{displaymath}
\sigma_b^{od} ({\rm 10~km}) ~=~ 2.2 \times 10^3~
\left ( {f \over 30 ~{\rm MHz}} \right )^{-1.45}
\end{displaymath}
\begin{equation}
\times
\left ( {E \over 10^{20}{~\rm eV}} \right )^{0.44}
\left ( {R \over 10 ~{\rm km}} \right )~{\rm ~m}.
\end{equation}

It is clear from this analysis\footnote{Here the linear increase of the 
cross-section with shower impact
parameter $R$ is due to the fact that the size of the first
Fresnel zone grows with distance. In fact, although we have
used the length of the Fresnel zone $L_F$ as the characteristic length 
of the ``thin wire'' portion of the shower, the actual length
of this region is likely to be much longer than $L_F$.}
that overdense showers will show
very large cross sections when the reflection is nearly
specular. However, as seen in Fig. 6, the cross sections decrease rapidly for
non-specular reflections. This behavior will be seen to contrast
sharply with the case of underdense scattering, to be treated
below. Any consideration of the detectability of a particular
shower must account for both contributions to the cross section.

In the analysis above we have also ignored the additional contribution due to
the partially coherent scattering of the much larger volume of
underdense plasma outside of $r_c$. This region could have 
refractive effects on the incoming radiation. In an appendix we
treat this issue in some detail and show that it produces
a negligible effect for the showers under consideration here.

\subsection{High-frequency/underdense case}

When either the frequency is high enough or the ionization density
low enough that there is effectively no region of the column where
the critical density obtains, the radar return is due to Thomson
scattering of the free electrons in the column, modulated by the
phase factor associated with the physical extent of the ionized
region as described in equation \ref{fft-eq} above. 

We have made
numerical estimates of the effective cross section for a restricted
set of cases here, with the methods 
described in more detail in an appendix. Here
we summarize the results.

\subsubsection{Normal incidence}

For the case of radar observations at normal incidence to the
shower, the near constancy of
the ionization density along the track allows us to estimate
the frequency dependence of the phase factor via a Fast Fourier
transform (FFT) of the two dimensional ionization distribution
in a plane transverse to the track. We have taken advantage
of this feature to make numerical estimates of the frequency-dependent
behavior of the phase factors at normal incidence.

Note that the requirement of normal incidence for the radar
rays interrogating the shower is not as restrictive as it might seem.
For example, consider a horizontal shower whose maximum is at a range of
$R_1$ from the radar transmitter/receiver system, and whose
axis is inclined an angle $\theta$ with respect to the ray
that joins the radar source and the shower maximum. Under these
conditions, the radar system can still interrogate the shower
along a ray that intersects the shower at normal incidence at
a distance $R_1 \cos \theta$ from the shower maximum. 

Thus a horizontal shower at 10 km altitude, with 15 km range to its
maximum, and inclined at $\theta = 60^\circ$ 
will have a ray that 
intersects it at normal incidence at a longitudinal 
distance along the shower of 7.5 km 
from shower maximum, at a range of 13 km, where the ionization line density
is still $\sim 60$\% of its maximum value. Thus the corresponding
acceptance solid angle over which these results apply is not
negligible, if the beam of the radar system is not too restrictive
in solid angle.

Fig. \ref{phasefac-fig} shows our estimates of the frequency dependence of the
phase factor for two horizontal
showers at altitudes of 5 and 10 km, for $E_0=10^{20}$ eV.
The phase factor depends strongly
on both frequency and shower altitude. The frequency dependence is
due to loss of coherence as noted above, and the altitude
dependence arises from  the change in Moliere radius to smaller
values at lower altitudes.
We have also investigated the behavior at other energies in this
regime and found these results to be insensitive to energy over
the range $10^{18}-10^{21}$ eV. For the two cases shown we have included 
fitted power laws 
\begin{displaymath}
\Phi(f; {\rm 5~km}) = 0.114~f_{10}^{-1.15}
\end{displaymath}
\begin{displaymath}
\Phi(f; {\rm 10~km}) = 1.13 \times 10^{-2}~f_{10}^{-1.84}~.
\end{displaymath}
Here $f_{10}$ is the radar frequency in units of 10 MHz.
These relations are
individually accurate to about 
20\% over the range from 10 MHz to 0.3 GHz.

Given the power loss associated with the phase factor,
the total cross section is then estimated by summing the
cross sections of the individual volume elements over one
Fresnel zone of the track, by analogy to the overdense case,
and multiplying by the overall phase factor for the radar frequency
and shower altitude.

Evaluating this modified cross section
for its energy and frequency dependence, we derive an empirical
expression for the case of a horizontal EAS at altitude of 10 km
\begin{displaymath}
\sigma_b^{ud} ({\rm 10~km}) ~=~ 175~
\left ( {f \over 30 ~{\rm MHz}} \right )^{-1.84}
\end{displaymath}
\begin{equation}
\times
\left ( {E \over 10^{20}{~\rm eV}} \right )^{1.9}
\left ( {R \over 10 ~{\rm km}} \right )~{\rm ~m^2}.
\end{equation}
and similarly for an altitude of 5 km:
\begin{displaymath}
\sigma_b^{ud} ({\rm 5~km}) ~=~ 1400~
\left ( {f \over 30 ~{\rm MHz}} \right )^{-1.15}
\end{displaymath}
\begin{equation}
\times
\left ( {E \over 10^{20}{~\rm eV}} \right )^{1.9}
\left ( {R \over 10 ~{\rm km}} \right )~{\rm ~m^2}.
\end{equation}

These relations should be valid out to ranges of several tens of km.
At the highest energies and lowest frequencies the underdense case may
not obtain, but these estimates will then provide a lower limit on
the RCS.

\subsubsection{Oblique angles}

For shower angles away from normal incidence, it is necessary to
consider all of the individual scattering contributions of the
electrons over the entire volume of the shower, and to treat
the pulse echo behavior explicitly. We discuss this in detail
in an appendix. We have numerically estimated the integrated
RCS for several wavelengths at a distance of
10 km to display some of the characteristics of the angular
dependence. 

\placefigure{thetfac10}
\begin{figure}[t]
\plotone{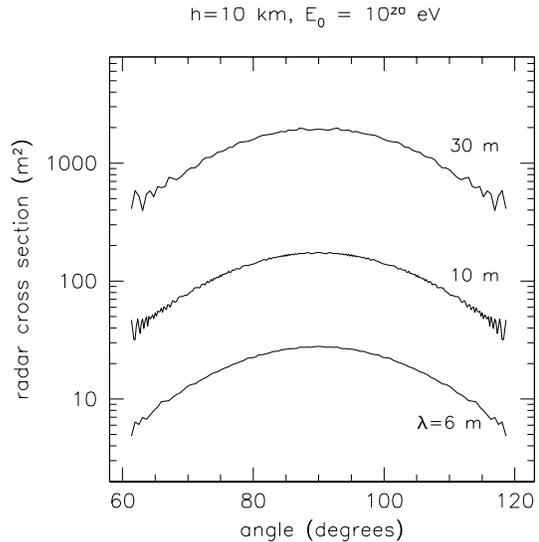}
\caption{Radar cross section over a range of angles centered
on normal incidence (at $90^{\circ}$) for a
horizontal air shower with
$E_0=10^{20}$ eV, at an altitude of 10 km, at radar wavelengths
30, 10, and 6 m. Results are 
based on numerically integrating the contributions of all of
the individual volume elements of the shower. \label{thetfac10-fig}}
\end{figure}

\placefigure{thetfac5}
\begin{figure}[t]
\plotone{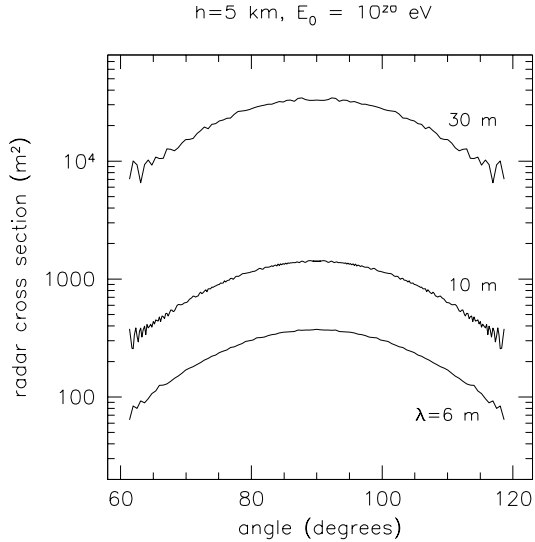}
\caption{Similar to previous figure, with
shower altitude of 5 km.\label{thetfac5-fig}}
\end{figure}

The results of this are shown in Figures \ref{thetfac10-fig} and 
\ref{thetfac5-fig}.
We have plotted the total RCS as a function of
angle over about 1 radian around normal incidence, for a 
horizontal shower at heights 10  and 5 km and energy $10^{20}$ eV,
and several different radar wavelengths.
The dependence on angle is quite different from the overdense
case where specular reflection is important; here, the cross
section has an approximately Gaussian profile with angle, with
a FWHM of order $30^{\circ}$. The fairly steep
wavelength dependence is also independent of angle over this
range. Thus the empirical relations derived in the
previous section can be used with appropriate factors for
the angle of incidence shown here.

We caution here that the angular dependence of the 
RCS at distances that are much greater than the length
of the shower is probably not accurately reflected by these
results. We have not yet derived a general analytical treatment
of the behavior for large distances, but we note that the
methods described in the appendix allow for numerical estimates
to be made at any distance.

\subsection{Caveats}

We note that, due to self-imposed limitations of scope,
a number of effects have not been treated in our
analysis. We list several of these here.

\paragraph{Geomagnetic effects.}
We have neglected geomagnetic charge separation in the shower
development. Although this is an important effect in determining
the detailed shower structure, it depends greatly on the shower
direction with respect to the magnetic field. Its effect
on the RCS would also depend on the angle of
observation and the treatment of this complex interaction is
beyond our scope at present.

\paragraph{Landau-Pomeranchuk-Migdal (LPM) effect.}
We have neglected the LPM effect in the showers considered here.
This effect, which retards the initial growth of the shower,
is important at the energies discussed and should be accounted for
in a more detailed treatment. However, it affects primarily
shower which interact initially through the electromagnetic rather
than the hadronic channel, and thus there are still many EAS at these
energies for which the approximation used is still accurate.

\paragraph{Secondary Fresnel zone echoes.}
For simplicity we have only treated the echo from the primary Fresnel
zone here. In fact, for the high SNR case, many distinct echoes may
be detected from successive Fresnel zones, and these will certainly
add significantly to the understanding of any detected shower.
On the other hand, for underdense showers that are not observed
near normal incidence, the time spread of the return will also
tend to decrease the instantaneous SNR. Such effects are important 
in understanding the response of a radar system to the wide
variety of possible EAS directions and relative angles and will need
to be explored further in any real application.

\paragraph{Near vertical EAS.}

For convenience we have analyzed a horizontal air shower, but such
showers are of general interest, since such highly-inclined showers imply
very deep initial interactions and are thus a possible signature of 
neutrino primaries. For more typical showers with small zenith angles,
the Moliere radius changes significantly during the shower 
development and complicated the analysis considerably.
Generalization of our analysis to EAS at all
angles is best done with a full numerical simulation. 
Preliminary results of such a simulation (D. Bergman 2000, pers. comm.) 
do not show significant differences in the ionization profiles,
and still show a tight core in the radial ionization density.

\paragraph{Pulse smearing by the shower.}

Note that our results for the underdense case above
apply in a general way only to non-pulsed radar,
since the superposition implied by the Fourier transforms assumes that
each Fourier component extends spatially out to dimensions that are
much larger than the shower column. For pulses whose compressed
time scale is physically shorter than the time-projected thickness
of the ionization column of a shower, we must explicitly account for
the behavior of the individual pulse echoes from different
portions of the shower. However, the effective RCS
corresponding to the integrated power of the echo is still described
by our analysis to first order. 

A general consideration of the complications
arising from the pulsed nature of the radar are beyond our scope
at present but must be included in any detailed estimate of the
pulsed radar echo behavior. In the appendix we describe a
method developed to deal with pulsed-radar reflections, and we 
utilize this method in some later results.

\paragraph{Plasma resonance effects.}
We have treated the underdense case in a simplified way which neglects
the possibility of plasma resonance effects. Such effects are in fact
seen in meteor echoes (Poulter \& Baggaley 1977)
and can play an important role in enhancing the
RCS, and in producing more complex time structure in
the radar return.

\section{Discussion}

In the previous sections we have presented evidence that EAS ionization 
columns are well within the range of VHF radar detection
for cascade energies above $\sim 10^{18}$ eV. Now we turn to a discussion
of the applicability of this approach to present efforts at detection
and characterization of such EAS. To establish a basis for comparison,
we first outline the capabilities of fluorescence detector systems,
to which EAS radar detection, if possible, would be most similar.

\paragraph{Fluorescence detector capabilities.}
The most capable present fluorescence detector system is the
High Resolution Fly's Eye system in Utah (Abu-Zayyad et al. 1999)
operating for several years now in ``monocular'' mode, and
now beginning to come on line with stereo-mode operation,
using a pair of telescope clusters separated by 12.6 km.
HiRes uses about 1 photomultiplier
per square degree of focal-plane sky coverage, with 2-3 m aperture 
low-resolution optical
telescopes used as the light collectors.

HiRes can detect events out
to 20-25 km, although at this distance the 
effects of aerosols increase the uncertainty of the measurements.
To achieve reasonable precision in the measurement of 
6 shower parameters (three geometric and three associated with
the energy and shower development) requires typically a minimum of
about 20 photomultiplier pixel hits in monocular mode, 
giving at this level
an energy resolution of $\sim 25$\% and precision of
about 30 gm cm$^{-2}$ in the position of shower maximum
(P. Sokolsky 2000, pers. comm.). 
Significant improvement in the precision of these
measurements is expected once the stereo mode
comes into full operation (Abu-Zayid et al. 1999). 
Observation efficiency for any Fluorescence detector
is limited to about 10\% since the observations can
only be made on clear, moonless nights. The total acceptance
aperture for HiRes is expected to be about 6000 km$^2$ sr in the
region of $10^{20}$ eV shower energy.

\subsection{Standalone EAS radar system.}

Here we consider what type of radar system would be required to
produce EAS measurements that can be compared to those
expected from HiRes. 

Because of the issues of interference rejection, 
a standalone EAS radar system would almost certainly
have to be constructed with 
multiple stations configured to both transmit and
receive, with pulse encoding such that each station would be
sensitive to pulses produced by the other stations. Bistatic
radar cross sections of EAS will vary only modestly from the
monostatic RCS values treated here. Global synchronization of the
stations to several tens of ns or less is routine with the 
Global Positioning System, particularly with the disabling of 
Selective Availability (SA) of the high precision GPS codes.

To optimize the signal-to-noise ratio of the detected echoes,
we wish to operate at the lowest frequency that is practical from
considerations of background noise, including ionospheric noise
from daytime operation. Although lower frequency operation
may be possible, we choose here to consider only frequencies
above about 30 MHz. In the HF-VHF regime the effective system noise
temperature for a remote site (distant from urban noise sources)
can be written as (Skolnik 1986):
\begin{equation}
T_{sys} ~=~ 2.9 \times 10^6 \left ( {f \over 3~{\rm MHz}}
\right )^{-2.9}~~{\rm K}
\end{equation}
which yields $T_{sys} = 3600$K at 30 MHz. 

Assuming that each station receives an echo from its own as well as 
all others in an array of $M$ transmit/receive stations, there are
$M+M(M-1)/2$ complex measurements for each combined pulse of the
$M$ stations. For six EAS parameters to be
fitted, the minimum number of stations required is $M=3$
(giving 6 total geometric ranges and echo amplitudes), 
assuming that the measurements are independent. Given 
that there will be some covariance of the measurements, and requiring that
the fit be well-constrained, at least 4 or 5 stations will be
needed, giving 12-20 ranges and amplitudes. Additional
closure-delay (or triangle) measurements number $M(M-1)(M-2)/6$, 
and will provide an extra 4-10 independent 
constraints for the $M=4,~M=5$ cases, respectively. However,
these latter constraints apply only to the geometry, which is
typically well-constrained in any case, and not the
echo amplitude. Addition of dual-polarization capability, which
is not uncommon in radar systems, would double the number of
measurements.

\subsubsection{Radar data types.}
The measured data can be characterized as follows:

\paragraph{Range data \& shower angle.} The time delay to the
leading edge of the echo at each station
will provide the range from the transmitter to the receiver.
The ranges can be measured typically to a precision of
roughly the compressed pulse width divided by the SNR.
For a SNR=10, and a compressed pulse width of 250 ns, 
the range resolution is below 10 m. 

A complete determination of the angular resolution possible
using radar is beyond our scope, but simple arguments can show
that a set of 4 or 5 stations should provide reasonable 
estimates of the track direction.
A schematic view of the geometry associated with the
range measurements for a pair of stations is shown in Fig.
\ref{angres-fig}. Here the stations 1 and 2 measure
leading-edge ranges $R_1$ and $R_2$ to separate portions of the shower
at along-track positions $s_1,s_2$ with respect to shower maximum.
The stations are separated 
by a baseline $B_{12}$ which is not generally parallel
to the shower track. Let a vector ${\bf n}$ bisect the baseline
and the shower segment from $s_1$ to $s_2$. The shower 
direction can then be characterized by angles $\alpha$, which
measures the rotation of the shower track around the axis 
${\bf n}$ with respect to the projected baseline, and $\beta$ which
measures the angle between the shower and ${\bf n}$.

\placefigure{angres-fig}
\begin{figure}[t]
\label{angres-fig}
\plotone{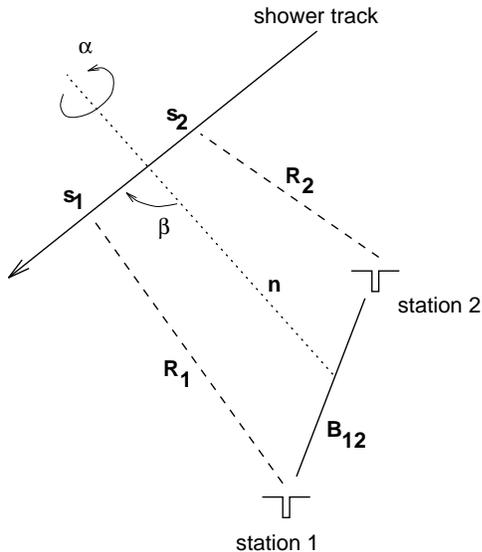}
\caption{A schematic of the range constraints on shower angles
provided by a pair of radar stations 1 and 2, separated
by baseline $B_{12}$. See text for details.\label{angres-fig}}
\end{figure}

It is evident from Fig. \ref{angres-fig} that changes in $\beta$
will under most conditions produce first order changes in
the range values, eg. 
\begin{equation}
\sigma_{\beta} ~\sim~ {\sigma_R \over | s_2 - s_1 |}
\end{equation}
where $\sigma_R$ is the mean range error and $\sigma_{\beta}$ the
resulting angular resolution in $\beta$. Since for tracks
that are primarily transverse to the baseline $B_{12}$, we
have $| s_2 - s_1 | \sim B_{12}$, 
then baselines of order 5 km or more with range errors
of $\sim 10$ m will give 
angular resolution in $\beta$ of order several mrad,
or about $0.1^{\circ}$.

Changes in $\alpha$ for a pair of stations under typical
expected detection geometry do not give first order effects 
in range, and are thus more problematic. Because a change in
$\alpha$ corresponds to a kind of ``torsion'' mode of
the range contraints (using a rigid-rod analogy), the
error in $\alpha$ for a two station solution is likely to
generally behave as
\begin{equation}
\sigma_{\alpha} ~\sim~ {\sqrt{2R~\sigma_R} \over | s_2 - s_1 | }
\end{equation}
which applies most directly to the case where $\alpha \ll 1$. 
For example, for $R= 20$ km, $\Delta s = 5$ km, and $\sigma_R = 10$ m,
we have $\sigma_{\alpha} \simeq 7^{\circ}$, much lower precision
than can be achieved in air fluorescence detectors. We note however 
that the addition of more radar stations, with the inclusion of
bistatic ranges and multiple
polarization measurements should rapidly improve the
precision as more constraints are added, but further estimates
are beyond our scope at present.

\paragraph{Echo amplitude data.} The echo amplitude is more
difficult to interpret. For underdense showers, the
echo amplitude is directly proportional to the number of electrons
present, and thus is similar to fluorescence detectors
in terms of its ability to estimate the number of ions
within a certain region of the track. However, since the
radar echo may be spread out over a band of different ranges,
determination of the contributing portion of the shower for 
each portion of the returned echo at each receiving station
is a clearly more complicated problem than that faced by
fluorescence detectors, which rely on imaging precision to
solve this problem. However, since the geometric parameters
of the shower track are easily separable via the range data,
the resulting amplitude of the leading edge of the return
can be attributed unambiguously to the portion of the
ionization column that is nearest to the receiving station.
By thus reconstructing a profile of these prompt received amplitudes
at their relative positions along the geometric track,
the profile of the shower development can be extracted.  

For overdense showers, the radar echos may show anomalously
high strength compared to expectations based on an underdense
echo. In this case, interpretation of the data may rely on
distinguishing these two cases on an event-by-event
basis. Estimates of the effective conductivity of the plasma
can then still yield the electron content of the shower.

In either of these cases, the amplitude data, although not
as straightforward to interpret at that of fluorescence
measurements, still provides the type of calorimetry 
of an EAS that is necessary to determine the shower energy.
This is a feature that particle counter arrays cannot
match in detail, and it sets apart the fluorescence detection
and potential radar detection techniques from other methods.

\subsubsection{Duty Cycle.}

Perhaps the most interesting feature of radar as a potential
EAS detection approach is the possibility of much higher duty cycle
observations over detection volumes comparable to those
of fluorescence detectors. In principle a radar system
is not limited to night operation, and suffers very little
attenuation from clouds or aerosols. Although operation in the
presence of lightning within the detection volume might
produce some spurious events, avoiding operation in periods of lightning
would cost at most a few percent of the livetime. 

The largest
uncertainty in efficiency is probably due to the free electron
lifetime, since if this is very short, it may
be difficult to ensure adequate pulse repetition frequency to
detect all showers. And although radar propagation is
unaffected by aerosols and clouds, the free electron lifetime
is not immune to variations in these components of the atmosphere,
and in practice they may be important limitations to the efficiency.
In addition, the presence of the sun above the horizon will increase
the system temperature somewhat, especially near solar maximum, and
this will act to increase the detection threshold for daytime 
operation.

\subsection{SNR Budget for a standalone EAS radar system}
\begin{figure}[t]
\label{array1}
\plotone{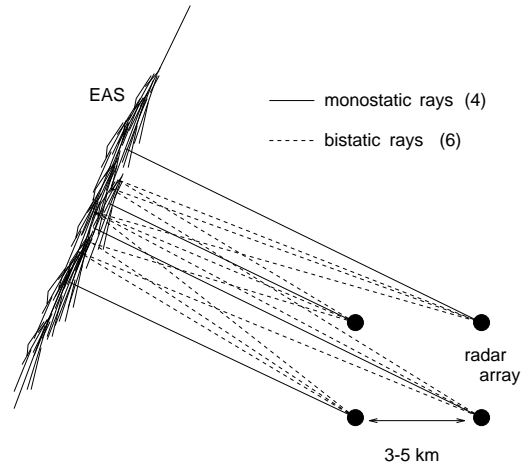}
\caption{An array geometry for a possible standalone EAS radar
detection system. Each of the 4 stations has both transmit \& receive
capability, and can distinguish the signals from the other stations
as well.\label{array-fig}}
\end{figure}

In Figure \ref{array-fig} we show a plan view of the geometry of a
radar array using both monostatic and bistatic detection of
an EAS, here shown with 4 stations at 3-5 km spacing, 
providing 10 ranges and
10 pulse amplitudes. 
Here we describe the parameters required of
each transmit/receive station to satisfy the
needs of the array described.

\placefigure{array1}

If we assume that we require an energy threshold of 
$10^{19}$ eV out to a range of 20 km, the implied
RCS for the underdense case 
is $\sim 2$ m$^{2}$ at this energy. The Fresnel zone length 
at this range is about 250 m for a shower 
at 10 km altitude, corresponding to about
10 g cm$^{-2}$ along the track. Using a commercial VHF radar system,
it is straightforward to
achieve 60 kW peak power and pulse repetition rates of 10-50 kHz
for $10~\mu$s pulses. The sky brightness temperature will lead to 
a system temperature of $T_{sys} = 3600$ K at 30 MHz as noted above. 
Since we wish to be sensitive to as large a volume as possible,
we use an antenna with a broad beam (thus corresponding to a low
gain). A suitable system is a vertical monopole or discone system
(Balanis 1997) which provides good $2\pi$
azimuthal coverage from $5^{\circ}$
to about $50^{\circ}$ elevation, giving an overall gain of $\sim 3$ with
respect to an isotropic antenna. We also assume here a somewhat 
pessimistic overall transmit/receive efficiency of $\eta = 0.05$,
and a single pulse echo (no pulse averaging)
to account for possible rapid decay of
the free electron population.

Using these assumptions we show the SNR budget, in dBm 
(decibels referenced
to 1 milliwatt) in Table \ref{auger-tab}. 
This budget corresponds to a single monostatic or bistatic 
pulse measurement in the array.
The resulting SNR of 6.4 provides a range 
resolution of 34 m at a 30 MHz operating frequency, using a chirp bandwidth
of 10\% (3 MHz in this case). 

{\footnotesize
\begin{table}[t]
\tablenum{1}
\centering
\caption{SNR budget for one station of a possible EAS standalone radar system.}
\label{auger-tab}
\medskip
\centering
\begin{tabular}{lcc}
\tableline
Parameter & value & $\pm$dBm \\ \tableline \\
{\underline{\it Received Power}} 
& $\sigma_b\eta P_t G^2\lambda^2 $&  \\
& $~~~~~~\times (4\pi)^{-3} R^{-4}$&  \\ \\
Peak transmit power &  60 kW & 77.8   \\
Pulse duration & $10\mu$s & ...    \\
chirp bandwidth & 3 MHz & ...  \\
Number of repetitions & 1 & 0.0 \\
Antenna gain & 3 & 9.54  \\
wavelength & 10 m & 20.0  \\
$\sigma_b$ at $E= 10^{19}$ eV & 3.8 m$^2$ & 5.8   \\
range to EAS & 20 km &  -172.0\\
xmit/rcv efficiency & 0.05 &  -13.0 \\
$(4\pi)^{-3}$ & $5.04 \times 10^{-4}$ &  -33.0 \\ \tableline \\
{\bf Received power} & & -104.9 dBm \\ \tableline \\
{\underline{\it Noise power }} & $k T_{sys} \Delta f$& \\ \\
Boltmann's constant &$1.38\times 10^{-20}$& -198.6 \\
	& ${\rm mW K^{-1} Hz^{-1}}$&   \\
System temperature & 3650 K & 35.6  \\
effective bandwidth & 100 kHz & 50.0 \\  \tableline \\
{\bf Noise power} & & -113.0 dBm \\ \tableline \\
{\bf SNR } & 6.4 &  8.0 dB \\
\tableline \\
\end{tabular}
\end{table}
}

We can also estimate the rate of shower detection as a function of
energy. To do this we estimate that the acceptance solid angle of 
the system is $\sim 1$ sr, based on the RCS within $10^{\circ}$ of
the peak values. This is significantly lower than that of
a fluorescence detector, and reflects the rapid decrease in the
RCS as the angle of incidence moves away from the normal to the
shower axis. We also consider only the RCS for showers at
10 km altitude, although showers at lower altitudes will have
significantly higher cross sections in general. The integral 
number of events seen
above a given energy $E_{thr}$ is given by 
\begin{equation}
Q(E_{thr}) ~\simeq~ \pi R_{eas}^2~ \eta_{obs} ~T~ \Omega~ I(E\geq E_{thr}) 
\end{equation}
where $\eta_{obs}$ is the observing efficiency, $T$ the observation duration, $\Omega\simeq 1$ is the effective solid angle, and
\begin{equation}
I(E\geq E_{thr}) \simeq 
\left ({E_{thr} \over 10^{19}~{\rm eV}}\right )^{-2} 
{\rm km^{-2} sr^{-1} yr^{-1}} 
\end{equation}
is the integral cosmic ray spectrum at
these energies. Here the maximum range for detection is
determined by inverting equation ~\ref{sn-eq} above, which yields:
\begin{displaymath}
R_{eas} ({\rm km}) ~=~ 82~\left ({S/N \over 10}\right )^{-1/3}
	\left ({E_{thr} \over 10^{20}~{\rm eV}}\right )^{0.63}
\end{displaymath}
\begin{displaymath}
	\times \left ({P \over 1~{\rm kW}}\right )^{1/3}
	\left ({\eta \over 0.1}\right )^{1/3}
	\left ({G \over 10}\right )^{2/3}
\end{displaymath}
\begin{equation}
	\times \left ({T_{sys} \over 1000~{\rm K}}\right )^{-1/3}
	\left ({\Delta t \over 10 ~\mu{\rm s}}\right )^{1/3}~.
\end{equation}

This relation is approximately valid to energies of order
$10^{20}$ eV, where the detection range is $\sim 80$ km.
Beyond this range, the minimum beam
elevation is such that it exceeds the initial
assumption of 10 km shower altitude, and the detection efficiency
as a function of shower altitude must be more explicity addressed
However, this exercise highlights the large potential detection
area for a radar system

\begin{figure}[t]
\label{easrate}
\plotone{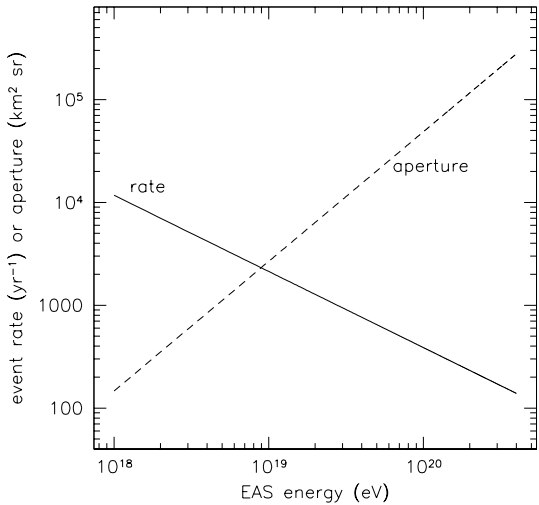}
\caption{The expected event rate detected at a SNR=6 for the
standalone EAS radar system described in the text and table 1.
The effective aperture corresponding to this rate is also plotted.
\label{easrate-fig}}
\end{figure}

\placefigure{easrate}

In Figure \ref{easrate-fig}, we plot the expected event rate
per year for 80\% observing efficiency, which is achievable with
a fully operational system. Here the assumed minimum $SNR=6.0$. Also
plotted with respect to the right axis is the effective aperture,
in km$^2$ sr, for the same assumptions.

It is evident from this analysis that EAS detection by radar 
certainly merits further investigation as a high-duty cycle
alternative to existing EAS detection systems. Although
it is not a competitive technique at lower energies, it 
appears that this approach may have significant benefits for
detection of EAS at the highest energies. Although it may not
be able to compete with fluorescence detection in
its ultimate accuracy in the estimation of shower parameters,
a system which utilizes a trigger provided by an existing particle
detector array such as the Auger Observatory could potentially gain
valuable complementary information on daytime events where
fluorescence information cannot be obtained.

\section{Conclusions}

We have demonstrated that, using standard models for the average behavior
of extensive air shower development, the resulting ionization appears
straightforward to detect using radar techniques in the VHF
frequency range (30--100 MHz), for primary energies greater than
about $10^{18}$ eV. We estimate that a relatively
modest ground-based system could provide information on EAS that is 
comparable to that available from other techniques, and may
do so with a much higher duty cycle and potentially a larger
effective aperture. The approach requires no new technology, making
use of standard radar equipment and detection techniques, and may
thus also provide a low-cost alternative to existing methods.

\acknowledgements
We thank an anonymous referee who provided a very valuable critique.
We thank Douglas Bergman, George Resch, Pierre Sokolsky, Paul Sommers, David Saltzberg, and Trevor Weekes for useful discussion and
comments. This research has been performed at the Jet Propulsion
Laboratory, California institute of Technology, under contract with
the National Aeronautics and Space Administration.

\appendix

\section{Appendix 1: Numerical Evaluation of RCS for the underdense case}

To estimate the RCS 
for the underdense regime of an EAS ionization column, we need
to evaluate the individual contributions of the scattering electrons
and the phase of the resulting scattered fields. As noted in
section 2 above, under certain conditions this is
equivalent to determining the Fourier transform of the electron
distribution. Here we describe the basis for this conclusion and
provide details of the methods used for numerical evaluation of the
RCS under these circumstances.

\subsection{Geometry of the problem}

\placefigure{radargeom}
\begin{figure}[t]
\plotone{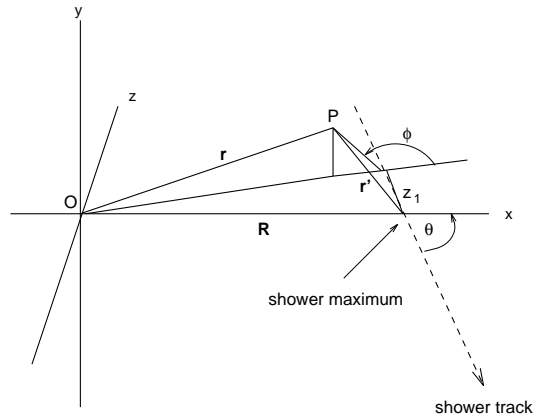}
\caption{The geometry for numerical evaluation of the summed radar
cross section of an ionization column for the case of oblique angle
of incidence. The radar transmitter/receiver is at the origin $O$, and
the angle of the track is $\theta$ with respect to the direction from the
origin to shower maximum $R$. The vector to the position $P$ in the 
ionization column that is being evaluated is ${\bf r}$, and the
vector separation of $P$ from the position of shower maximum is ${\bf r'}$.
$\phi$ is the azimuthal angle of $P$ in cylindrical coordinates 
with respect to the shower axis. \label{geom-fig}}
\end{figure}

The geometry for this evaluation is shown in Fig. \ref{geom-fig}.
Here we let ${\bf R}$ represent the vector distance from the 
origin $O$ (at the radar transmitter/receiver) to the position
of shower maximum. 
We choose the $x-y$ plane to be coincident with the plane
containing the axis of the shower and ${\bf R}$.
Within the ionization column itself, we use
cylindrical coordinates $\rho,\phi,z_1$
to calculate the proper ionization density at
position $P$ with respect to the position of shower maximum; 
${\bf r'}=(\rho,\phi,z_1)$ is the vector separation of $P$ from shower maximum.
The vector joining the origin to $P$ is denoted as ${\bf r}$. 
The angle between the shower axis and ${\bf R}$ is denoted $\theta$.

Given these definitions, we find
\begin{equation}
|{\bf r} |^2 = \rho^2+z_1^2+R^2 + 2R(\rho\cos\phi\sin\theta-z_1\cos\theta)
\end{equation}
which expresses the distance to an arbitrary point in the ionization
column as a function of the distance to shower maximum, the local
cylindrical coordinates of the point, and the angle of obliquity $\theta$
of the shower.

\subsection{Scattered field}

The electric field that is incident at $P$ can be written as
\begin{equation}
U_{inc} ~=~ U_{tr} G_{tr} e^{i\omega t} 
{e^{i{\bf k \cdot r}} \over |{\bf r}|}
\end{equation}
where $U_{tr}$ is the magnitude of the transmitted field, $G_{tr}$ is the
transmitter gain factor (as defined in the main text), $k=2\pi/\lambda$,
and ${\bf r}$ is defined in Fig. \ref{geom-fig}. Assuming that ${\bf k}$
and ${\bf r}$ are parallel (that is, neglecting the effective dielectric
behavior of the underdense plasma in the column), the received field
from a small volume of electrons $n_e dV$ which scatter coherently
is then given by
\begin{equation}
dU_{rcv} ~=~ U_{inc} \sqrt{\sigma_T} n_e dV
G_{sc} e^{i\omega t} {e^{i{\bf k}_{sc} \cdot {\bf r}_{sc}} 
\over |{\bf r}_{sc}|}
\end{equation}
where $G_{sc}$ is the effective gain of the electron scattering process
($G_{sc}=1$ for isotropic scattering),\footnote{For Thomson scattering,
$G_{sc}\simeq 1.6$ corresponding to electric dipole emission. However,
this response is complicated in any real plasma by a number of effects,
and we have assumed $G_{sc}=1$ for all cases considered here.}
and ${\bf k}_{sc},~{\bf r}_{sc}$ 
are the wavevector and radial vector of the scattered wave. In this case we
have ${\bf k}_{sc} = -{\bf k}$ and ${\bf r}_{sc} = -{\bf r}$, and thus
\begin{equation}
dU_{rcv} ~=~ U_{tr} G_{tr} G_{sc} \sqrt{\sigma_T} n_e dV e^{i\omega t} 
{e^{2 i{\bf k \cdot r}} \over |{\bf r}|^2}
\end{equation}
which displays the expected radial dependence of the field (equivalent
to a $r^{-4}$ dependence in received power), and also the additional factor
of two in the spatial phase which is due to the two-way trip of the radiation.

If we assume for the moment that the time variation of the field is
steady--state, we can drop the time dependence, and integrate over the
volume of the shower to yield the total received field strength,
assuming that the transmitted beam is larger than the dimensions of the
shower. If we also assume that the distance from the shower is large
enough that $|{\bf r}| \approx |{\bf R}|$, then we have
\begin{equation}
\label{fft0-eq}
R^2 U_{rcv} ~=~ U_{tr} G_{tr} G_{sc} \sqrt{\sigma_T}
\int_V n_e({\bf r}) e^{ i{\bf q \cdot r}} d^3{\bf r}
\end{equation}
which shows (taking ${\bf q} = 2{\bf k}$) that the received field is
proportional to the Fourier transform of the electron density
distribution. 

\subsection{Normal incidence}

Under conditions where the pulse is incident on the ionization
column at normal incidence, and if we assume that the ionization
column is essentially constant along its longitudinal length, the
received field becomes proportional to the two--dimensional Fourier
transform of the cross--sectional electron density, times the
length of the Fresnel zone $L_F$ (defined as the length over which the
scattered fields are in phase):
\begin{equation}
\label{fft1-eq}
R^2 U_{rcv} |_{\theta=0}~=~ U_{tr} L_F G_{tr} G_{sc} \sqrt{\sigma_T}
\int_A n_e({\bf r}) e^{ i{\bf q \cdot r}} d^2{\bf r}~.
\end{equation}
We have numerically evaluated this using Fast Fourier transform
techniques, since the area $A$ over which this is done is of
order $A=\pi r_m^2$ and the number of $\lambda_{min}/8$ cells in the grid is
$64 A/\lambda_{min}^2$, where we have assumed $\lambda_{min}\sim .5$ m,
giving a $(2^{13})^2$ grid. The result of the FFT can be 
normalized to provide a frequency--dependent phase factor, which we
have plotted in Fig. \ref{phasefac-fig} above.

\subsection{Pulsed radar \& the time--projected RCS}

In the previous analysis we have assumed a steady--state incident field 
(which corresponds to carrier--wave or CW radar). 
However, in our analysis of
section 4, we assumed that the radar is pulsed, and in fact pulsed
radar is preferable to achieve the kind of time resolution required in
this application. For the overdense case, the analysis presented
in section 4 is adequate, since we are considering reflections from
a localized source at the critical radius of the ionization column.
Also, for normal incidence, the reflections from the first
Fresnel zone arrive closely in phase and can be treated without
explicitly accounting for the pulse echo behavior.

When we allow for oblique shower angles, however, the concept of
the Fresnel zone loses its usefulness, and
the received echo comes from the entire volume of the
shower and is thus significantly spread out in time. We require a
method of treating the behavior of a discrete pulse under these
conditions.

\begin{figure}[t]
\plotone{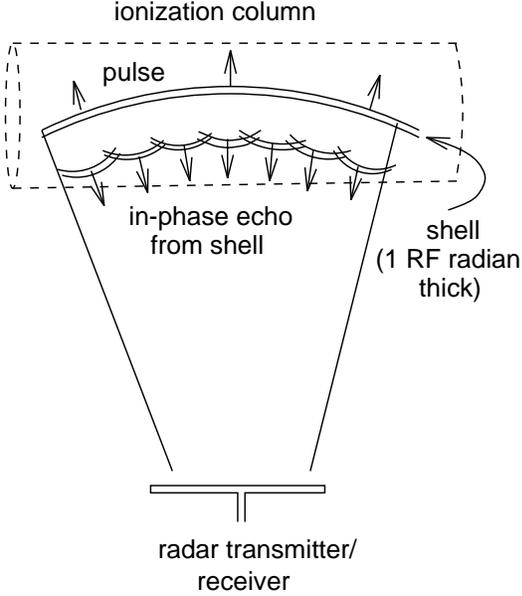}
\caption{Illustration of the geometry that gives
rise to the concept of the time-projected radar
cross section. The ionization column can be divided into
radial shells such that all of the electrons that scatter
the incoming radar are in phase (to within a radian) within the
shell. \label{tprogeom-fig}}
\end{figure}

The approach we have followed is intuitively simple: we first produce
a time-projected profile of the electron density; that is, we
calculate the round-trip time associated with the global phase of each of the
contributing differential volume elements of the scattering electrons
\begin{equation}
t_i ~=~ {\lambda \over 2 \pi c} ~{\bf q \cdot r}_i~.
\end{equation}
Here $t_i$ is the round-trip time for the $i$th volume element at
location ${\bf r}_i$. If we consider all of the volume elements for
which $t_i = t_j$ within a specified interval $\Delta t \leq \omega^{-1}$, 
then we define a shell at distance $r = t/2c$ with a thickness of
1 radian of phase at the radar angular frequency $\omega$. 
This is depicted in Fig.~\ref{tprogeom-fig}.

\placefigure{tprogeom}

Summing all of the contributing elements in this shell gives the
time--projected square root of the total cross section at that phase:
\begin{equation}
\label{y-eq}
Y (t) \Delta t ~=~  \sqrt{\sigma_T}  \int_{r-t/4c}^{r+t/4c} 
n_e({\bf r}) d^3{\bf r}~.
\end{equation}

Once $Y(t)$ is determined for a given radar wavelength,
shower angle and energy, and distance, then the pulse response of the
ionization column is determined by convolving the electric
field of the pulse profile with $Y(t)$, and squaring the
result. If the pulse is defined to have unit amplitude, then the
resulting convolution can be expressed as an effective radar cross
section for use in equation \ref{sn-eq}.

For example, if the pulse is assumed to be a $\delta$-function in time,
the resulting convolution $\Upsilon(t)$ is
\begin{equation}
\Upsilon (t) ~=~ \int_{-\infty}^{\infty} Y(t')\delta(t-t') dt' = Y(t)
\end{equation}
and the received power is then proportional to $|Y(t)|^2$. For
other pulse profiles we numerically evaluate the convolution explicitly,
and take the maximum of the resulting profile to be the effective radar
cross section.

\placefigure{timepro}
\begin{figure}[t]
\plotone{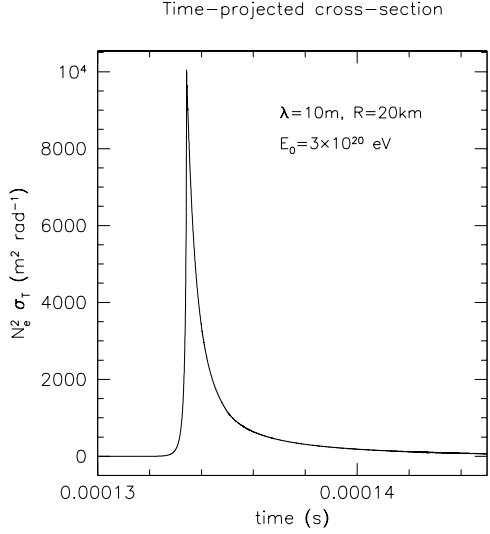}
\caption{ The square of the function $Y(t)$, 
which gives the time--projected total radar
cross section, for an ionization column produced by a 
shower with $E_0=3 \times 10^{20}$ eV, for  
$\lambda = 10$ m, at a distance of 20 km. The time
axis is windowed around the arrival time of the round trip
travel time. $Y(t)$ may be thought
of as a kind of Green's function time response of the ionization column
to a sharp radar pulse, with a magnitude that is proportional
to the square root of the RCS at a given pulse phase.
 \label{timepro-fig}}
\end{figure}

In Fig. \ref{timepro-fig} we show an example of the
square of $Y(t)$, giving the time--projected 
cross section calculated according to the manner described 
here. In each case the time bins correspond to a radian of angular phase
according to the wavelength indicated, and the value for each bin is 
estimated according to equation \ref{y-eq}. The profile in this
case shows a characteristic rapid rise on its leading edge, reflecting
the rise in cross--sectional ionization density of the nearest
portion of the shower. The more slowly falling trailing edge is 
due to the contributions from portions of the shower further away.

We note that $Y(t)$ defined in this way is analogous to a Green's
function for the time response of the system. It can be
used in this manner to evaluate arbitrary radar pulse shapes,
for a given EAS energy and direction. For extremely
broad-band radar systems, the effective RCS values will begin to
approach those of $Y(t)$. For more narrow-band radar pulses,
the convolution of the field oscillations with $Y(t)$ will
act to partially cancel and the effective RCS will decrease.

\section{Appendix 2: Refractive effects of the plasma halo}

The refractive index of a plasma is given by Spitzer (1962)
in the quasilongitudinal approximation by
\begin{equation}
\label{refrac-eq}
n ~=~ \left [ 1 - \left ( {\nu_p \over \nu }\right ) \right ]^{1/2}
\end{equation}
where we have assumed that the frequency of the radiation is
well above the electron gyrofrequency in the earth's magnetic
field ($\sim 840$ kHz). In this case the phase delay of the
radiation after propagating a distance $l$ through the plasma
is given as (cf. Sovers et al. 1998):
\begin{equation}
\Delta_{pd} ~=~ -q/\nu^2
\end{equation}
where the variable $q$ here represents the column density of
the electrons integrated along the ray path:
\begin{equation}
q ~=~ { cr_0 \over 2 \pi } \int n_e dl
\end{equation}
where $r_0$ is the classical electron radius and $c$ is the speed of light. 
The negative phase delay, or phase advance given above is characteristic of
propagation in a plasma. More appropriate for consideration here is
the group delay, obtained by differentiating $\phi = \nu \Delta_{pd}$
with respect to frequency
\begin{equation}
\Delta_{gd} ~=~ q/\nu^2
\end{equation}
which constitutes a net additive delay that increases at lower 
frequencies. Numerically, this delay has the value
\begin{equation}
\Delta_{gd} ~=~ 1.34 \times 10^{-3}~ \chi_e / \nu^2 ~{\rm sec}
\end{equation}
where $\chi_e$ is the electron column density in e$^-$ cm$^{-2}$

Integrating the electron densities in the
halo of a $10^{20}$ eV shower along a radial path from outside the
shower down to a radius of 1 m and then back out
again (to approximate the radar ray path) yields
$\chi_e \simeq 4 \times 10^8$ e$^-$ cm$^{-2}$, and the corresponding
delay at 30 MHz is 0.6 ns, or about $6^{\circ}$ of RF phase. A group delay 
of this order will have a negligible effect on the received power
compared to the geometric phase factors. At lower frequencies,
such at 10 MHz, the group delay is an order of magnitude larger,
but the induced RF phase lag is still only $\sim 20^{\circ}$.
Thus it appears that the refractive effects of the radial plasma
halo can be ignored to first order for the EAS ionization
columns considered here. 

Similar analysis of ray-path deviation 
along chords in the outer halo (due to
the gradient in the plasma density) shows that the refraction is
not significant until the chords approach within several
m of the EAS core for the highest energy showers. Thus although
it is important to include such effects in any
final calibration of parameters estimated by radar
measurements, they do not appear to bear significantly on the
detectability of the showers.

\end{document}